\documentclass[usenatbib]{mn2e}
\usepackage{float}
\usepackage[pdftex]{graphicx}
\usepackage[utf8]{inputenc}
\usepackage{rotating}
\usepackage{comment}

\renewcommand{\vec}[1]{ {\bmath #1} }

\def\gsim{ \lower .75ex \hbox{$\sim$} \llap{\raise .27ex \hbox{$>$}} }
\def\lsim{ \lower .75ex \hbox{$\sim$} \llap{\raise .27ex \hbox{$<$}} }

\newcommand{\sigmaSIDM}{\sigma_{\rm SIDM}}
\newcommand{\offset}{\Delta {\bf s}}
\newcommand{\SIDM}{{\rm SIDM}}
\newcommand{\Var}{\mathrm{Var}}

\usepackage{ifthen}
\newboolean{thesis}
\setboolean{thesis}{false}

\usepackage{lscape}
\usepackage[justification=centering]{caption}

\usepackage{afterpage}

\usepackage{amssymb}
\usepackage{amsmath}
\usepackage[amssymb]{SIunits} 

\usepackage{times}

\usepackage{paralist}

\usepackage[bookmarks,bookmarksnumbered,colorlinks=true,citecolor=blue,linkcolor=blue]{hyperref}
\usepackage{color}

\usepackage{pbox}

\usepackage{booktabs}

\usepackage[flushleft]{threeparttable}

\usepackage{comment} 

\definecolor{darkgreen}{rgb}{0.0, 0.5, 0.0}

\usepackage{booktabs}
\usepackage{hhline}
\usepackage{breqn}
\usepackage{standalone}
\usepackage{dcolumn}
	\newcolumntype{d}[1]{D{.}{.}{#1}}
\usepackage{tabularx}
\usepackage{booktabs}
\usepackage{microtype}
\usepackage[]{algorithm2e}
\usepackage[update,prepend]{epstopdf}
\graphicspath{{figures/finalized/}}

\def\Hmat{\mathbf{\it H}}
\def\apjl{ApJL }

\def\apjs{ApJS }

\def\aap{A\&A }

\providecommand{\adsurl}[1]{\href{#1}{ADS}}
 
\title[	Galaxy-DM offsets in Illustris clusters]
{Offsets between member galaxies and dark matter in clusters:\\ 
a test with the Illustris simulation}

\author[Karen Y. Ng et al.]{Karen Y. Ng,$^1$
	Annalisa Pillepich,$^2$$^,$$^4$ 
	David Wittman,$^1$
	William A. Dawson,$^3$ 
	\newauthor Lars Hernquist,$^{2}$
	Dylan R. Nelson$^{2}$$^,$$^5$ \\ 
$^{1}$Department of Physics, University of California Davis, One Shields
Avenue, Davis, CA 95616, USA\\ 
$^{2}$Harvard-Smithsonian Center for Astrophysics, 60 Garden Street, Cambridge,
MA 02138, USA\\
$^{3}$Lawrence Livermore National Laboratory, P.O. Box 808, Livermore, CA 94551-0808, USA \\
$^{4}$Max-Planck-Institut f{\"u}r Astronomie, K{\"o}nigstuhl 17, 69117 Heidelberg, Germany\\
$^{5}$Max-Planck-Institut f{\"u}r Astrophysik, Karl-Schwarzschild-Str. 1, 85741 Garching, Germany\\}

\begin{document}
\pagerange{\pageref{firstpage}--\pageref{lastpage}}
\pubyear{2015} \maketitle\label{firstpage}
\begin{abstract} 
	Dark matter with a non-zero self-interacting cross section ($\sigma_{\rm
	SIDM}$) has been posited
	as a solution to a number of outstanding astrophysical mysteries.
	Many studies of merging galaxy clusters have given constraints on
	$\sigma_{\rm SIDM}$ based on the spatial offset between the member galaxy
	population and the dark matter distribution.      
	Assuming $\sigma_{\rm SIDM} = 0$, how likely is
	it for us to see the galaxy-DM offset values observed in merging clusters of galaxies? 
	To answer this question, we formulate a hypothesis test using data from Illustris, a $\Lambda$CDM cosmological 
	simulation.
	We select 43 Illustris clusters and their galaxy members at $z\sim0$ by mimicking observations; 
	we quantify their level of {\it relaxedness}; and examine the accuracies of commonly used galaxy summary 
	statistics, including kernel-density-estimation (KDE) luminosity peak, KDE number density peak, shrinking aperture, centroid and
	the location of the brightest cluster galaxy (BCG), with broad applications to optical studies of
	galaxy clusters.
	We use the dark-matter particles from the simulation to reproduce commonly adopted methods to identify dark-matter peaks based on gravitational lensing cluster maps. By analysing each cluster in 768 projections, we determine the optimistic noise floor in the measurements of the galaxy-DM offsets. We find that the choice of the adopted galaxy summary statistics affects the inferred offset
	values substantially, with the BCG and the luminosity peak giving the tightest
	68-th percentile offset levels, $\lesssim$ 4 kpc and $\lesssim$ 32 kpc, respectively. We also find a long
	tail of the offset distribution of the BCG due to projected 
	substructures. Galaxy summary statistics such as
	shrinking aperture, number density and centroid give a large offset scatter of $\sim$ 50
	$-$ 100 kpc at	the 68-th percentile level, even for clusters with only one dominant 
	mass component.
	Out of the 15 reported offsets from observed merging clusters
	that we examined, 13 of them are consistent with Illustris unrelaxed cluster offsets at the 2-sigma (95-th percentile) level, i.e. consistent with the hypothesis that $\Lambda$CDM is the true underlying physical model. 
	
\end{abstract}


\section{Introduction} 

During the latest stages of structure formation, the universe has given birth to
non-linear, hierarchical structures known as galaxy clusters. 
These clusters, made up of dark matter, galaxies and hot gas,
are constantly accreting mass, merging with other gravitationally-collapsed objects and evolving with their
environments. Together with their member galaxies, they trace the underlying dark matter (DM) 
distribution, highlighting its overdensities. 
Long recognized as powerful cosmological probes, galaxy clusters are the ideal sites where to test assumptions regarding the nature of DM: 
in their dense regions, the rates of particle
interactions may be enhanced, including the hypothetical self-interaction of DM
particles (hereafter, SIDM). SIDM with a small cross section
$\sigmaSIDM$, instead of a Cold-Dark Matter (CDM) model with zero $\sigma_{\rm
SIDM}$, has been invoked as a potential solution to a few apparent discrepancies between our current cosmological model and observations, e.g. the 
``core vs cusp" problem in the central regions of
dwarf galaxies \citep[see e.g.][and references therein]{Vogelsberger:2012b, Rocha:2013a, Vogelsberger:2014c}.\\

The study of more massive systems like galaxy clusters provides an alternative test bed for SIDM theories.
In particular, efforts have focused on constraining SIDM observables with merging galaxy clusters.
As the SIDM scattering rates can be shown to depend on the 
relative velocities of the DM particles \citep{Markevitch2004}, the high
velocities and column densities in cluster mergers may produce a SIDM signal. This idea was verified by 
\cite{Randall2008d}, who used the first suites of simulations of the
Bullet Cluster that included SIDM physics. \cite{Randall2008d} showed that the scattering events of SIDM can cause the DM
to lag behind the relatively collisionless galaxies, thus resulting in a spatial {\it offset} between the DM component and the cluster galaxy population:
\begin{equation}
	\offset_{\SIDM} \equiv {\bf s}_{\rm gal} - {\bf s}_{\rm DM},
	\label{eq:offset}
\end{equation}
where $\bf{s}_{gal}$ and $\bf{s}_{DM}$ are the two-dimensional (2D) spatial
locations of the summary statistics of the galaxy population and the summary statistics of the DM density (usually its peak), respectively. 
Interestingly, \cite{Randall2008d} also showed an almost linear dependence of
$\offset_{\SIDM}$ with $\sigmaSIDM$ in simulations of merging clusters.
By comparing the simulated offsets to the observed one in the Bullet Cluster (\citealt[25 $\pm$ 29 kpc][]{Markevitch2004} and \citealt{Bradac2006b}),
\cite{Randall2008d} were able to derive a constraint of $\sigmaSIDM <$ 
1.25 cm$^2$ g$^{-1}$.  \\

These early results have motivated a burst of activity and the development of so-called staged simulations of merging clusters in SIDM scenarios (\citealt{Randall2008d}, \citealt{Kahlhoefer14}, \citealt{Robertson2016}, \citealt{Kim:2016}).
On the observational side, the list of merging galaxy 
clusters studied to constrain SIDM has been growing. 
The majority of these studies reported non-zero,
but statistically {\it in}significant offsets, including the Musketball cluster
\citep{Dawson2013}, MACSJ1752 \citep{Jee2015}, 
and others that we list and comment in more detail in Table \ref{tab:offset_results}.
With observationally motivated 
levels of $ \sigmaSIDM < 3 \centi\meter^2 / \gram$, different simulations have consistently reported the maximum SIDM offset 
signals ($\lesssim 50$ kpc) to be a few times smaller than the observed offsets 
($ \gtrsim 100$ kpc).  When \cite{Kahlhoefer14} simulated SIDM with both low-momentum-transfer 
self-interaction and rare self-interactions of DM with high momentum transfer, they found maximum 
offsets that are $< 30$ kpc for $\sigmaSIDM$ as high as 1.6 \centi\meter$^2$ / \gram.
The reported offset from \cite{Randall2008d} for $\sigmaSIDM = 1.25~\centi\meter^2 / \gram$ is slightly larger: 53.9 kpc. 
Other newer simulations also reach similar conclusions.
\cite{Kim:2016} find a maximum offset $< 50$ kpc for 
$\sigmaSIDM = 3~\centi\meter^2 / \gram$, while
\cite{Robertson2016} find a maximum offset $\lesssim 40$ kpc  
 from a simulation suite of a Bullet Cluster analog 
 with $\sigmaSIDM = 1~\centi\meter^2 /$ \gram.
 
To match the observed offset values of $\sim 100$ kpc with the one predicted by staged SIDM simulations, it is certainly possible in theoretical experiments
to increase $\sigmaSIDM$ further. However, large $\sigmaSIDM$ values are not favoured, as they would cause a large rate of halo evaporation \citep{Kim:2016}, which has not been seen in observations. 
In conclusions, the discrepancies between the maximum offset values inferred by simulations and those from observations point towards other possible contributions to the
observed offsets: would the offsets shrink with better data or is the uncertainty mostly intrinsic? 

While staged simulations provide suitable settings for understanding the physical origin and statistical distribution of $\offset_{\SIDM}$
for merging galaxy clusters, the simulated offsets should not be interpreted a priori as the observed ones. 
If we assume (statistical) independence of 
the different possible contributions of the observed 
offset, we can decompose the model of $\Delta s_{\rm obs}$ to be:
\begin{equation}
	\offset_{\rm obs} = \offset_{\SIDM} + n + \cdots,
	\label{eq:signal_and_noise}
\end{equation}
with $\offset_{\SIDM}$ being the offset distribution caused by SIDM, $n$ being the distribution of observational uncertainties, systematic bias and statistical noise, and ``$\cdots$" denoting any contribution from unknown physical causes.
So far, staged merger simulations have given estimates of the distribution of the $\offset_{\SIDM}$ term by adopting, by construction,
a highly suppressed $n$ term. However, statistical and observational uncertainties might be significant. Galaxies are
sparse samples of the underlying DM overdensities --- it is possible that the 
summary statistics of the sparse sample are different from those of the 
underlying distribution.
  
In fact, constraining SIDM with merging clusters via eq. \ref{eq:offset} is further complicated by the fact that there is no theoretical
foundation showing which observable is the most sensitive to each
possible type of SIDM. Typically, the offset is inferred by first determining ${\bf s}_{\rm DM}$ 
and ${\bf s}_{\rm gal}$ independently before taking their difference. While the measurements of ${\bf s}_{\rm DM}$ are usually based on finding the peak of gravitational lensing maps, there is no standard and univocal procedure for summarising the sparse member galaxy distribution.
For example, \cite{Kahlhoefer14} have argued that SIDM 
does not cause significant offsets between the galaxy and DM {\it peaks}, but only leads to an offset
between the corresponding {\it centroids} within the dynamical timescale for
relaxation ($\sim$ several Gyr). 

Staged merger simulations themselves are affected by a series of simplications and choices which minimize uncertainties contributions and systematics. We list the assumptions from the staged merger simulations as follows. First, the physical properties of the galaxy clusters are oversimplified e.g. both the DM and 
galaxies were initialized to follow a parametric spatial 
distribution   
(\citealt{Randall2008d}, \citealt{Kahlhoefer14}, \citealt{Robertson2016}, \citealt{Kim:2016}), 
such as a Navarro-Frenk-White (NFW) profile. For real observations,
substructures and foreground
contaminations can all make the inference of the spatial distribution 
of the galaxy population more uncertain. 
Second, at the beginning of the simulated mergers, the galaxy and the DM
population are set to have zero offsets for any $\sigmaSIDM$ value. 
This assures that the offsets obtained from the merger simulations are due to SIDM. 
While this initial condition 
is a reasonable choice for making the effects of SIDM stand out from the simulations, 
the real $\offset$ of a cluster does not have to be zero at the 
beginning of a merger. We call this the intrinsic offset $\offset_{\rm
intrinsic}$.
Third, these staged merger simulations commonly use a much higher number of 
galaxies than is observable. \cite{Randall2008d} used
$10^5$ galaxies, \cite{Kahlhoefer14} used as many galaxy proxies as DM particles, 
while \cite{Kim:2016} used either 5.7k or 57k galaxies. 
Fourth, the mergers are usually initialized with conditions that maximize SIDM
interaction rates, such as a zero impact parameter. However, the impact
parameter in any given observed merger is highly uncertain. 
Fifth, physical processes such as feedback from Active
Galactic Nuclei (AGN) are also sometimes ignored in these simulations. 
AGN feedback can affect the spatial distributions between the galaxy summary
statistic, such as  Brightest Cluster
Galaxy (BCG) and  
the DM peak \citep{Cui2015}. This effect can increase the percentage of large 
BCG offsets in the tail region of the distribution and can affect whether an observed offset can be 
considered as a statistically significant deviation from the CDM model based on the p-value.

 

In this paper, we begin to bridge the gap between observations and staged simulations of merging clusters by adopting a large scale hydrodynamical cosmological simulation, Illustris (\citealt{Vogelsberger2014, Vogelsberger:2014b, Genel:2014,Sijacki:2015}). 
The Illustris volume contains a sample of a few tens galaxy clusters {and groups, in a variety of evolutionary stages.
As it includes the effects of baryonic physics and follows the stellar populations based on the consistent evolution of the cosmic initial conditions, Illustris data provide a more realistic sample of galaxies than the ones that are prescribed onto (i.e. added after the fact by a semi-analytical model) DM-only cosmological simulations \citep[e.g.][]{Harvey2013d}. The high accuracy treatment of self-gravity provide realistic dynamics and produce a realistic distribution of cluster subhaloes \citep{Vogelsberger2014}. Since the profiles of the galaxy clusters were not input in any way, such as through symmetrical, parametric forms, we are sensitive to the natural asymmetries of cluster profiles which are inaccessible in analytical studies or staged simulations. Finally, since Illustris contains full information on the dark matter, stellar, and gaseous content of all clusters, robust comparisons to the observations can be made.

The Illustris simulation has been evolved in a $\Lambda$CDM cosmology, namely it assumes no SIDM. In this paper, we therefore quantify the level of observational uncertainties, systematic biases, and statistical noise that arise in the measurement of the galaxy-DM offsets in clusters in a CDM scenario ($n_{\rm CDM}$). 
We desire to determine the {\it optimistic} noise floor by assuming perfect knowledge about cluster galaxy and dark matter membership (i.e., disregarding line of sight structure effects), and uncertainties in the dark matter peak identification smaller than the ones faced by lensing measurements. If the offsets in this optimistic study are of order the observed offsets or smaller, then it will be worth extending the study to include other sources of noise.
Additionally, we propose a hypothesis test to determine whether the observed offset data can be compatible with offsets derived from a CDM simulation. Since the Illustris simulation assumes no SIDM but includes other physical effects and statistical noise, 
our study is complementary to the staged simulations for understanding $n$ and $\offset_{\rm intrinsic}$.


In the following sections, we will 
1) extract realistic observables from the Illustris simulation for
comparison to observation; 2) explore the properties of a series of statistics commonly adopted to   
summarize the spatial distribution of the {\it the member galaxy population} in a galaxy cluster; and 3)	
give estimates for the offsets between the summary statistics of the galaxy  
population and the DM distribution in $\Lambda$CDM cosmology.
Then, we will 4) investigate what physical conditions or cluster properties produce extremes in the Illustris offset distributions; and 5) statistically compare the Illustris $\Lambda$CDM results with reported galaxy-DM offsets inferred from observations of merging clusters.

The organization of this \ifthenelse{\boolean{thesis}}{chapter}{paper} is as follows.
In Section \ref{sec:illustris_sim}, we describe the data from the Illustris simulation  
and the selection criteria and setup we have employed to ensure similarity with observational methods.
In Section \ref{sec:methods}, we define the various 
summary statistics of the spatial distribution of galaxies, describe how we prepare our dark
matter spatial data to resemble lensing convergence maps, and detail the way we measure the galaxy-DM offsets in the simulated projected clusters. In Section \ref{sec:validation} we show the statistical performance of the different summary statistics before we show the main results
in Section \ref{sec:results}. In Section \ref{sec:discussion}, 
we discuss the implications of our results and compare to other simulations and observations. We summarize and conclude in Section \ref{sec:conclusions}.

Our analysis makes use of the same flat Lambda Cold Dark Matter ($\Lambda$CDM) cosmology
as the Illustris simulation. The relevant cosmological parameters are
$\Omega_\Lambda = 0.7274, \Omega_m = 0.2726$, $H_0 = 70.4$
km~s$^{-1}$~Mpc$^{-1}$, and $\sigma_8 = 0.809$.\\

All the code used in this work is publicly available at the link: \href{https://github.com/karenyyng/galaxy\_DM\_offset}{https://github.com/karenyyng/galaxy\_DM\_offset}. There, a series of Jupyter Notebooks allows the retrieval of all Figures presented in this paper, including the visualisation for {\it all} selected Illustris clusters of maps and plots that, for brevity, below are shown just for a selection of cases. 

\begin{figure*}
	\includegraphics[width=\linewidth]{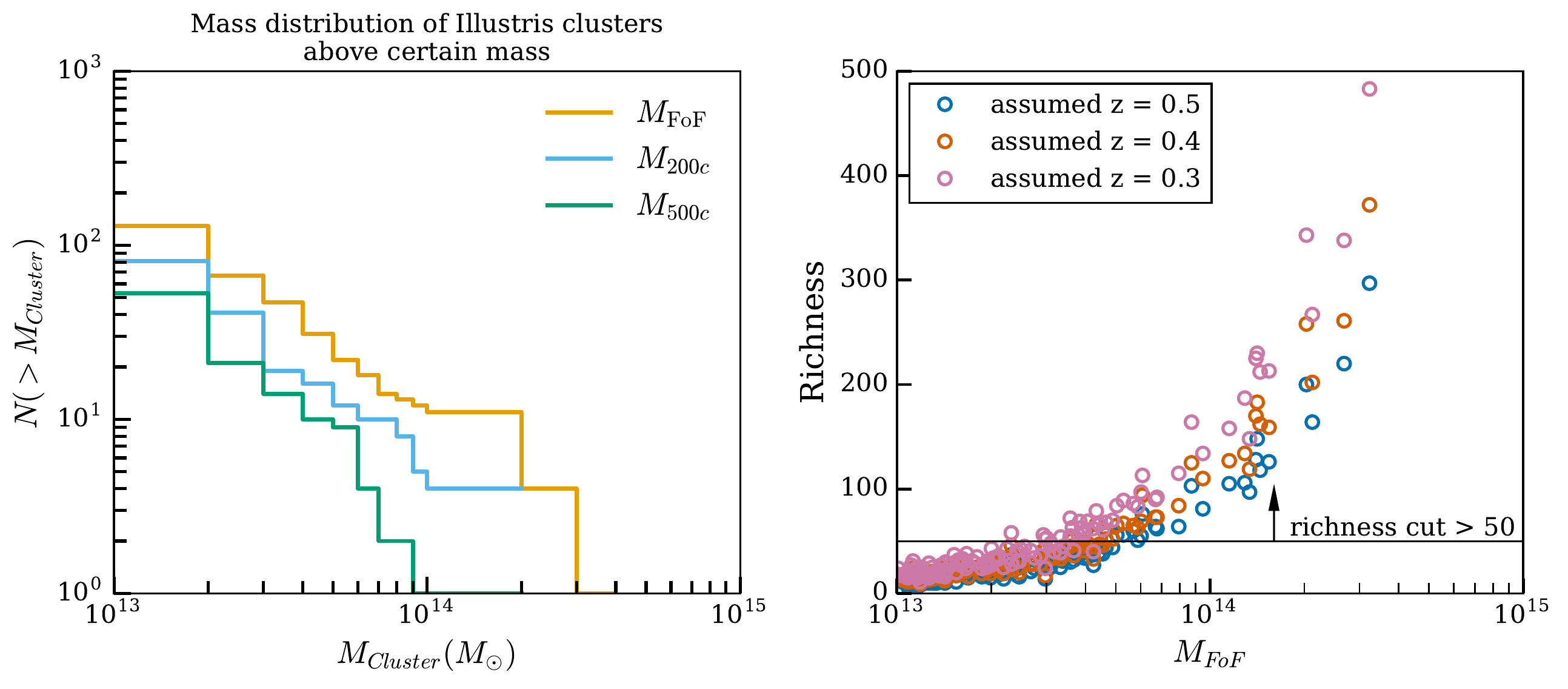}
	\caption{ {\bf Left:} Mass distribution of the group and cluster sized 
		DM haloes in Illustris at $z=0$, for different mass definitions. Mass estimates obtained by the
		FoF algorithm are labeled as  M$_{\text{FoF}}$.
		We use M$_{200c}$ and M$_{500c}$ to represent 
		masses that are centered on the most bound particle within $R_{200c}$ and
		$R_{500c}$ respectively. The  
		average densities within $R_{200c}$ and $R_{500c}$ are 
		200 or 500 times the critical density of the universe. 
		{\bf Right:} 
		Mass-richness relationship of llustris galaxy clusters and groups with 
		$M_{\rm FoF} > 10^{13} M_{\odot}$ assuming different cosmological redshifts
		of the observed clusters. Richness is defined as the number of galaxies within the FoF haloes and with $i \leq 24.4$ apparent at $z=0.3$.
\label{fig:mass_richness}}
\end{figure*}

\section{The Illustris simulation data} 
\label{sec:illustris_sim}

The Illustris volume \citep{Vogelsberger2014, Vogelsberger:2014b, Genel:2014,Sijacki:2015}  contains some of the most realistic and rich samples of simulated galaxy clusters to date. It follows the interplay of gravity, hydrodynamics and baryonic physics mechanisms on a uniform box of 75 comoving Mpc $h^{-1}$. 
Modelled with the moving-mesh code {\sc AREPO} \citep{Springel2010}, Illustris is characterised by a particle resolution of 1.6 and 6.3 $\times 10^6$ M$_\odot $ (stellar and DM elements, respectively). The softening length of the DM as well as stellar particles is 1.4 comoving kpc down to $z=1$, after which point the softening of the stars is fixed to its $z=1$ value of 0.7 physical kpc. The softening of the gas cells, which have a mean baryon mass of about $10^6$ M$_\odot $, is adaptive in space and time.

The Illustris simulation contains a sophisticated model for the key physical processes relevant for galaxy formation \citep{Vogelsberger:2013b, Torrey:2014a}. These include star formation, stellar population evolution with mass and metal production, tracking nine individual species from hydrogen to iron, and with a coupled primordial and metal line cooling implementation. Stellar feedback with a kinetic wind scheme generates galactic scale outflows, while feedback from AGN uses a dual state model, injecting thermal energy in the quasar mode and inflating hot bubbles in the radio mode, in addition to treating radiative proximity effects.

In addition to these astrophysical models, gas dynamical processes and environmental effects that shape the properties of the intra-cluster plasma and the galaxies residing therein, such as tidal and ram pressure stripping, dynamical friction and merging, are self consistently modeled by the hydrodynamical treatment. This makes Illustris especially suitable to verify observational methods pertaining to the spatial distribution of DM in clusters and of cluster galaxies.

In what follows, we focus on the data from snapshot number 135 (cosmological $z=0$) of the Illustris-1 simulation \citep{Nelson:2015b}. Among the different Illustris simulation implementations,
the Illustris-1 simulation (hereafter simply ``Illustris'') has the highest resolution and has demonstrated excellent to reasonable agreement with a broad number of observational scaling relations and galaxy properties at low redshift \citep{Vogelsberger:2014b,Sales:2015,Torrey:2015,Snyder:2015} as well as across cosmic time \citep{Genel:2014}.

\subsection{Selection of Illustris Clusters}

Gravitationally bound objects, including the clusters we study here, were identified by processing each Illustris snapshot using the SUBFIND algorithm \citep{Springel:2001, Dolag:2009}. This method first executes a standard friends-of-friends (FoF) group finder \citep{Davis1985}, within which gravitationally bound substructures are then located and characterized hierarchically. We refer to independent particle associations as \textit{haloes} (our clusters being the most massive haloes), each of which may contain one or more satellite galaxy, \textit{subhaloes}. These subhaloes may be either dark or luminous, and the luminous satellite galaxies of clusters are used to estimate our summary statistics.

The most massive halo in the Illustris box at $z=0$ has a total mass of about $2\times10^{14}$ M$_\odot$ (see Figure \ref{fig:mass_richness}, left panel). For the present analysis, we make use of clusters with a minimum richness of 50, namely with a minimum of 50 gravitationally bound satellite galaxies above a reasonable observational brightness limit, apparent $i \leq 24.4$, which is feasible for spectroscopic confirmation with the DEIMOS instrument on Keck at $z=0.3$.
For this, we use the absolute AB magnitudes extracted by \cite{Vogelsberger2014} for each luminous subhalo in
the SDSS bands of $g, r, i, z$ using stellar population synthesis models.
There is relatively large statistical uncertainty if we try
to analyze groups with less than 50 member galaxies. \\

As indicated by the right-hand panel of Fig. \ref{fig:mass_richness}, 
a total of 43 clusters has survived this magnitude cut. These simulated galaxy clusters (or groups) have 
masses ranging from about $10^{13}$ M$_\odot $ to $2 \times 10^{14}$ M$_\odot$.

\subsection{Cluster properties}
\label{subsec:cluster_properties}

Basic properties of the 43 selected Illustris galaxy clusters are reported in Table \ref{tab:cluster_prop}. 
They encompass a variety of evolutionary and relaxedness stages.

\subsubsection{On the dynamical states (relaxedness) of the galaxy clusters}
\label{subsubsec:relaxedness}

Clusters undergo merger activities at a large range of physical scales and 
on time scales of millions of years. Their dynamical history cannot be directly 
quantified from observations. 

\ifthenelse{\boolean{thesis}}{\begin{landscape}}{}
\begin{table*}
\begin{center}
	\caption{Summary of the selection criteria and setup for the adopted sample of Illustris galaxy clusters and their member galaxies.
\label{tab:member_galaxy_selections}} 

\begin{tabular}{@{}lcccc@{}}
\hline 
Data &  Selection strategy  & Sensitivity & Relevant section  \\ \hline
Field of view (FOV) & FoF halo finder& comparable to FOV of the Subaru
Suprime camera & \ref{sec:FOV}  \\ 
Observed filter & $i$-band & consistent among the redder $r, i, z$ bands &   
\ref{subsec:galaxy_properties}
\\ 
Cluster richness  & $i \leq 24.4$ and $z = 0.3$  & sensitive to
the assumed cosmological redshift of cluster and & \ref{sec:illustris_sim} \\ 
& & the assumed limiting magnitude of telescope &   \\
Two-dimensional projections & uniform HEALPix samples over a sphere &
discussed in the result section \ref{subsec:projections} & \ref{subsubsec:projections}\\  
\hline
\end{tabular}
\end{center} 
\end{table*}
\ifthenelse{\boolean{thesis}}{\end{landscape}}{}

Within the simulation and theoretical community, commonly-adopted relaxedness criteria are the following:
\begin{itemize}
	\item unrelaxedness$_0$: the ratio between the mass which is gravitationally-bound in subhaloes and the total mass
		of the galaxy cluster. The lower the ratio, the fewer substructures
		are present in the cluster. 
	\item unrelaxedness$_1$: the distance between the most bound particle and
		the center of mass, normalized by R$_{200c}$ (the three-dimensional (3D) radius in which the
		average density is 200 times the critical density of the universe).
		The smaller the distance, the lower the unrelaxedness. 
\end{itemize}

In Table \ref{tab:cluster_prop} we report measurements of the Illustris clusters' relaxedness based on the definitions above. In Section \ref{subsubsec:nu}, we will relate such simulation-motivated criteria to a more observation-oriented quantities. 

\subsection{Selection of the field-of-view}
\label{sec:FOV}

Throughout this paper, for each cluster, we consider all the gravitationally bound DM particles as well as satellite galaxies as identified by the {\sc FoF} and {\sc SUBFIND} algorithms. This means that the cluster haloes can assume any shape and morphology, possibly with multiple cluster subcomponents at various stages of merging \citep[e.g.][]{Lukic:2009}. Moreover, this does not a priori impose a spatial restriction, and cluster members can extend to distances much larger than the cluster nominal virial radius.

In practice, we select for every cluster in our sample all the DM and satellite galaxies which belong to the FoF-halo and are contained in a cube encompassing the whole FoF-halo without additional distance cut.
In fact, this unrestricted field of view ($\gtrsim 1$ Mpc per side) can be much larger than
the field of view reported from Hubble Space Telescope 
observations, e.g. $~200$ kpc per side \citep{Zitrin2012a}. 
Yet, assuming $z = 0.3$, 
the projected extent for most of the Illustris galaxy clusters and groups
fits inside the field of view of different instruments, such as the Subaru Suprime Camera,
which covers a physical area of $\sim 9$ Mpc $\times 7$ Mpc at z = 0.3.

Moreover, throughout our analysis we will ignore foreground / background structures and galaxies. Indeed, we are interested in quantifying the noise floor that will exist even if observers can perfectly eliminate foreground/background contamination. Modeling the contribution of foreground/background contamination is beyond the scope of this paper, as it is largely data dependent: our scope is to provide the most useful result applicable to a variety of situations, and hence we will not include estimates which depend on foreground and background structures and galaxies.


\subsubsection{Spatial Projections}
\label{subsubsec:projections}
In observations, clusters and their galaxy members are projected along one fixed line of sight. 
In this paper, we therefore project our simulated clusters onto a number of 2D orthographic projections.
As the actions of projecting the data and estimating the summary statistics are
non-commutative, we first project the data before estimating any projected 
observable. 

In order to choose the sample of projection orientations, we use {\sc HealPy},
which is a {\sc Python} wrapper for
{\sc{HEALPix}}\footnote{HEALPix is
currently hosted at http://healpix.sourceforge.net}
\citep{Gorski2005}. {\sc HealPy} gives different
lines-of-sight, each centered on a {\sc{HEALPix}}
pixel. Each pixel covers the same amount of area on a sphere. 
The number of projections that we employed is 768 for each cluster. With these
many projections, the offset distributions of each cluster start to converge to a
stable distribution. 
Even though there are at least 2 identical projections for each cluster due to
one possible line-of-sight from the front and one from the back, it does not
affect any summary statistic. We do not remove the duplication as it breaks
the rotational symmetry in the 2D plane when we compute the 2D population
distribution of offsets.  

\begin{figure}
	\centering
	\includegraphics[width=\linewidth]{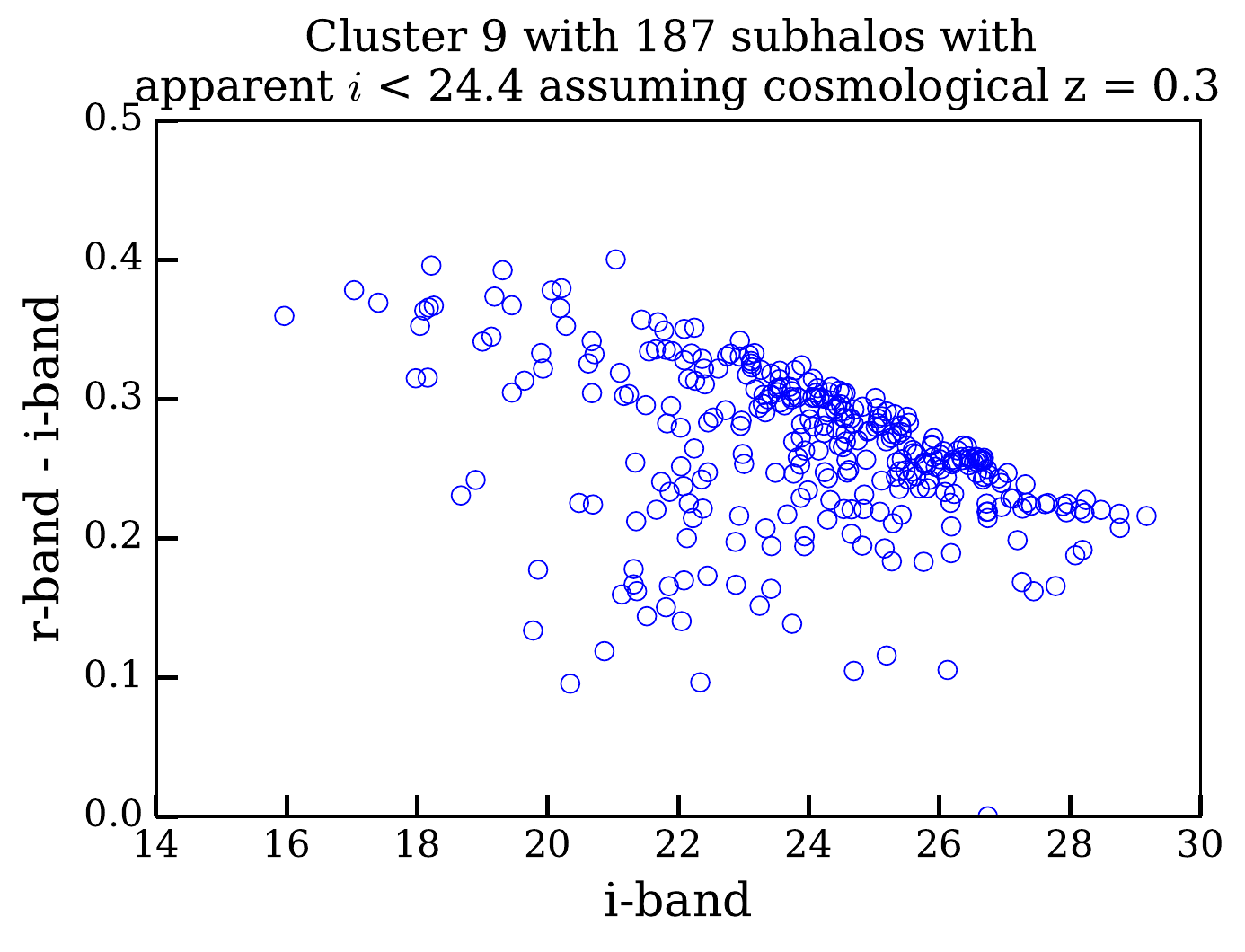}
	\caption{Color-magnitude diagram of one of the galaxy clusters selected from Illustris for analysis. This cluster is the 9th most massive, and is given as an example. 
		The apparent magnitude is calculated assuming that 
		the cosmological redshift is $z = 0.3$. 
		We can see a clear overdense region that corresponds to a red-sequence.
		Note that the lack of foreground / background galaxies results in a smaller
		range of colors than seen in most observed color-magnitude diagrams.	
		\label{fig:color_magnitude_diagram}
	} 
\end{figure}

\subsection{Properties of the galaxies in Illustris clusters}
\label{subsec:galaxy_properties}

Physical properties of Illustris galaxies are determined considering the stellar particles belonging to a given subhalo/galaxy according to the {\sc SUBFIND} halo finder. In the catalogs, Illustris galaxies are characterized by a plethora of properties, of which we mainly use the 3D position with respect to their cluster centre and the apparent magnitudes. 

Indeed, different galaxies have different masses and luminosities, so they should not be considered with equal
importance when e.g. considered for peak identification. One of the most common weighting schemes employed for galaxy data is based on luminosity in a particular band. For some of the methods, we will investigate
the differences in peak identification with and without any luminosity weights, and  we will show empirically that uniform
weights yield larger galaxy-DM discrepancies. Throughout, we will adopt the $i-$band magnitude
associated to each subhalo/galaxy as the weight. Since the $i-$band is
one of the redder bands, the mass-to-light ratio is not significantly biased due to star formation activity.

As a sanity check, we examine if the color distributions of galaxies in Illustris-1 are broadly
consistent with observations. Fig. \ref{fig:color_magnitude_diagram} shows the $r-i$ vs $i$ color-magnitude
diagram for the galaxy clusters in Illustris. It clearly reveals an overdense region of galaxies known as the red-sequence in the 
color-magnitude diagram. The red-sequence is prominent even if we
use other colors formed by different combinations of the $r, i, z$ bands.

\section{Methods}\label{sec:methods}

In this paper we quantify the galaxy-DM offset in simulated galaxy clusters for comparison to observations.
In particular, we compute the 2D distance between a) the {\it centre, peak} or summary statistics of the dark matter as traced by its smooth and clumpy components and observationally probed via e.g. gravitational lensing; and b) the {\it centre, peak} or summary statistics of the stellar distribution as traced by the discrete ensemble of cluster member galaxies and observationally probed with optical imaging.

In what follows, we will consider the Illustris clusters selected in Section \ref{sec:illustris_sim} and consider their galaxy members according to a set of choices and criteria summarized in Table \ref{tab:member_galaxy_selections}.

Observationally, a common way of summarizing the DM distribution in a
galaxy cluster is by finding the peaks of its gravitational lensing maps
(\citealt{Medezinski2013}, \citealt{Markevitch2004}, 
\citealt{Zitrin13}). In fact, the peak region is physically 
interesting due to the higher particle density and interaction rates and so particularly ideal to searches for SIDM.
The most direct analogous statistic for summarizing the member galaxy
population in a cluster is therefore also the peak, 
Comparing the DM peak with the summary statistics of the galaxy population that
are not the peak  can have an {\it offset} purely due to the difference in
the choice of the statistic. 

Here we define and adopt five commonly used 
point statistics or locations for summarizing 
the member galaxy population in a galaxy cluster:
1) location of the brightest cluster galaxy (BCG); 2) galaxies centroid; 3) shrinking aperture centroid; 4) kernel-density-estimation (KDE) based luminosity peak; 5) kernel-density-estimation (KDE) based number density peak.

We avoid any manual methods for
comparison purposes, as well as for scalability and reproducibility. 
Since all the methods listed in this \ifthenelse{\boolean{thesis}}{chapter}{paper}
are automated with the source code openly available, 
it is possible for future studies to reuse our code for comparisons. 
Another major advantage for automation is that it allows us  
to apply the same methods across the different snapshots of the (Illustris) simulations to
examine the variability of $\Delta s$ across time in future studies.

\subsection{Finding the galaxy summary statistic}
\label{subsec:methods:galaxy}

\subsubsection{Location of the Brightest Cluster Galaxies (BCG)}
The position of the BCG is a standard galaxy summary statistic.
The BCGs are formed by the merger of many smaller
galaxies. However, star formation can cause
less massive galaxies to be brighter in the bluer photometric bands.
To avoid star formation from biasing our algorithm for identifying the
BCG, we find the brightest galaxies in redder bands i.e. the $r, i, z$
bands: we find that they give consistent results for all selected clusters. 
Throughout this paper, we use the $i-$band to pick the BCG for computing the plots and the final results.

\subsubsection{Computing the weighted centroid}
\label{subsubsec:weighted_centroid}
The usual definition of the weighted centroid reads
\begin{equation}
	\bar{\bf x}_w = \frac{\sum_i w_i \vec{x}_i}{\sum_i w_i},
\end{equation}
with $\vec{x}_i$ being the positional vector of each member galaxy 
and where the weight $w_i$ is the $i$-band luminosity of the $i$-th galaxy
Centroids can be biased by subcomponents from merging activities. 
These estimates are also sensitive to odd boundaries 
of the field of view.

\subsubsection{Shrinking aperture estimates}

Another method among astronomers for finding the peak of a spatial
distribution is the shrinking aperture method, which iteratively
computes the centroid over successively smaller apertures.
We test if the shrinking aperture method is able to reliably recover the 
peak of the luminosity map.
This method is dependent on the initial diameter and the initial centre 
location of the aperture, and 
 does not evaluate if the cluster is made up of
several components.
The estimate using the shrinking aperture algorithm can be biased by
substructures. The only way to inform the algorithm about substructures would
be to introduce another parameter to restrict the extent of the aperture, or to
partition the data with another (statistical) algorithm.
More to the point, the convergence of results of this method is not guaranteed. We use a
convergence criterion of having the aperture distance not change more than 2\% 
between successive iterations as a reference. The actual implementation in
{\sc Python} can be found at the linked given in the Introduction
while
the pseudo-code can be found in Appendix \ref{app:shrink_apert}.

\subsubsection{Cross-validated Kernel Density Estimation (KDE) and the peak finder} 
\label{subsubsec:KDE}
Finding the exact peak of a set of data points 
involves computing the density estimate of the data points and sorting through
the density estimates. A specific version of this density estimation process is
known as histogramming. During the making of a histogram, each data point is
given some weight using a tophat kernel and the weights are summed up at
specific data locations (e.g. $\bf{x}_i$). 
A histogram is not good for peak estimate for {\it sparse} data for two reasons: 1) the
choice of laying down the bin boundaries affects the count in each bin, 2) the choice of
bin width also affects the count in the bin. Only when the available number of data points
for binning is large are the estimates of histograms and smoothed density
estimates approximately the same. The number density of Illustris cluster galaxies  
is sparse enough ($< 500$) for the uncertainty introduced by histogramming 
to bias our peak estimate. We therefore perform a kernel density estimation
instead of histogramming. 
For the density estimate of galaxy luminosity, 
we adopt a Gaussian kernel. 
The exact choice of the functional form of the smoothing kernel does
not dominate the density estimate as long as the chosen kernel is
smooth \citep{Feigelson2014}. 

For computing the density estimate, the most important parameter  
is the bandwidth of the smoothing kernel, 
which takes the form of a matrix in the 2D case. 
When the kernel width is
too large, the data is over-smoothed, 
resulting in a bias of the peak estimate. On the other hand, when the kernel
width is too small, it results in high variances of the estimate and 
too many peaks due to noise. 
The parameter decision for balancing between the portion of
bias or variance in an estimate is also known as the bias-variance tradeoff. 
A good illustration can be seen in \citealt{Vanderplas2012} from 
\href{http://www.astroml.org/book\_figures/chapter6/fig\_hist\_to\_kernel.html}{http://goo.gl/jvsfcv}.
All other smoothing procedures, including interpolation with splines,
polynomials, and filter convolutions, also face the same tradeoff. 

A well-known way to minimize the fitting error from the density estimate is through
a data-based approach called cross-validation to obtain 
the optimal 2D smoothing
bandwidth matrix ($\Hmat$) of the 2D Gaussian kernel for the
density estimate $\hat{f}$:
\begin{align}
	\hat{f}(\chi; \Hmat) &= \frac{1}{n} \frac{1}{(2\pi)^{d/2}|\Hmat|^{1/2}}
	\sum_{i=1}^n w_i \exp((\chi-{\bf x}_i)^T H^{-1} (\chi-{\bf x}_i)),
	\label{eq:cross_validated_bandwidth}
\end{align}
where the dimensionality is $d=2$ for our projected quantities,
$\chi$ represents the uniform grid points for evaluation, 
$\bf{x}_i$ contains the spatial coordinates for each of the identified member 
galaxies that survived our brightness cut, and $w_i$ is again a weight, i.e. a property of the galaxy. 
The idea behind cross-validation is to leave a small fraction of data points 
out as the test set, and use the rest of the data points as 
the training set for computing the estimated density.
Then it is possible to estimate and minimize the Asymptotic Mean-Integrated Squared Error
(AMISE)  by searching
for the best set of bandwidth matrix values, eliminating any free parameter. 

Specifically, we make use of the smoothed-cross validation \citep{Hall1992} 
bandwidth selector in the statistical package {\sc{ks}} \citep{Duong2007} 
in the {\sc{R}} statistical computing environment \citep{R_core}. 
Among all the different {\sc{R}} packages, {\sc{ks}} is the
only package capable of handling the magnitude weights of the data points 
while inferring the density estimates \citep{Deng2011}. 
Although this particular implementation of KDE has a computational runtime of $O(n^2)$, 
the number of cluster galaxies is
small enough for this method to finish quickly ($\lesssim 0.65$ second per
projection per cluster). 

The resulting KDE contains rich information about the spatial distribution of
the clusters, and we focus on the peak regions. We perform the cross-validated KDE in two fashions: a) by adopting as weights in Eq. \ref{eq:cross_validated_bandwidth} the $i-$band luminosity for each galaxy (leading to what we call ``KDE luminosity peaks''); and b) by adopting a constant weighting in Eq. \ref{eq:cross_validated_bandwidth} (leading to the identification of the ``KDE number-density peaks''.)
We employ both a first and second-order finite differencing algorithm to find the local maxima.  
The local maxima are then sorted according to the KDE density in a descending
fashion before we perform peak matching and compute the offset. The exact
procedure is discussed in section \ref{subsec:offsets}.

\subsection{Measure of the dynamical states via the KDE luminosity peaks}
\label{subsubsec:nu}
The KDE luminosity maps provide a tool to quantify the relaxedness of clusters.
For each projection of each cluster, we normalize the density of all 
luminosity peaks to those of the brightest peak. 
Luminosity peaks that sit on top of actual subclusters would then have a density 
comparable to those of the brightest peak. 
Then we sum the density of all the galaxy peaks for a cluster and call this value
$\nu$. When the value of $\nu$ is much bigger than 1, it indicates the presence 
of projected significant substructure(s). As $\nu$ is not expressed in terms of masses, it can be computed using galaxy
magnitudes from optical survey data. 

\begin{figure*}
	\begin{center}
	
	\ifthenelse{\NOT
		\boolean{thesis}}{\includegraphics[width=0.78\linewidth]{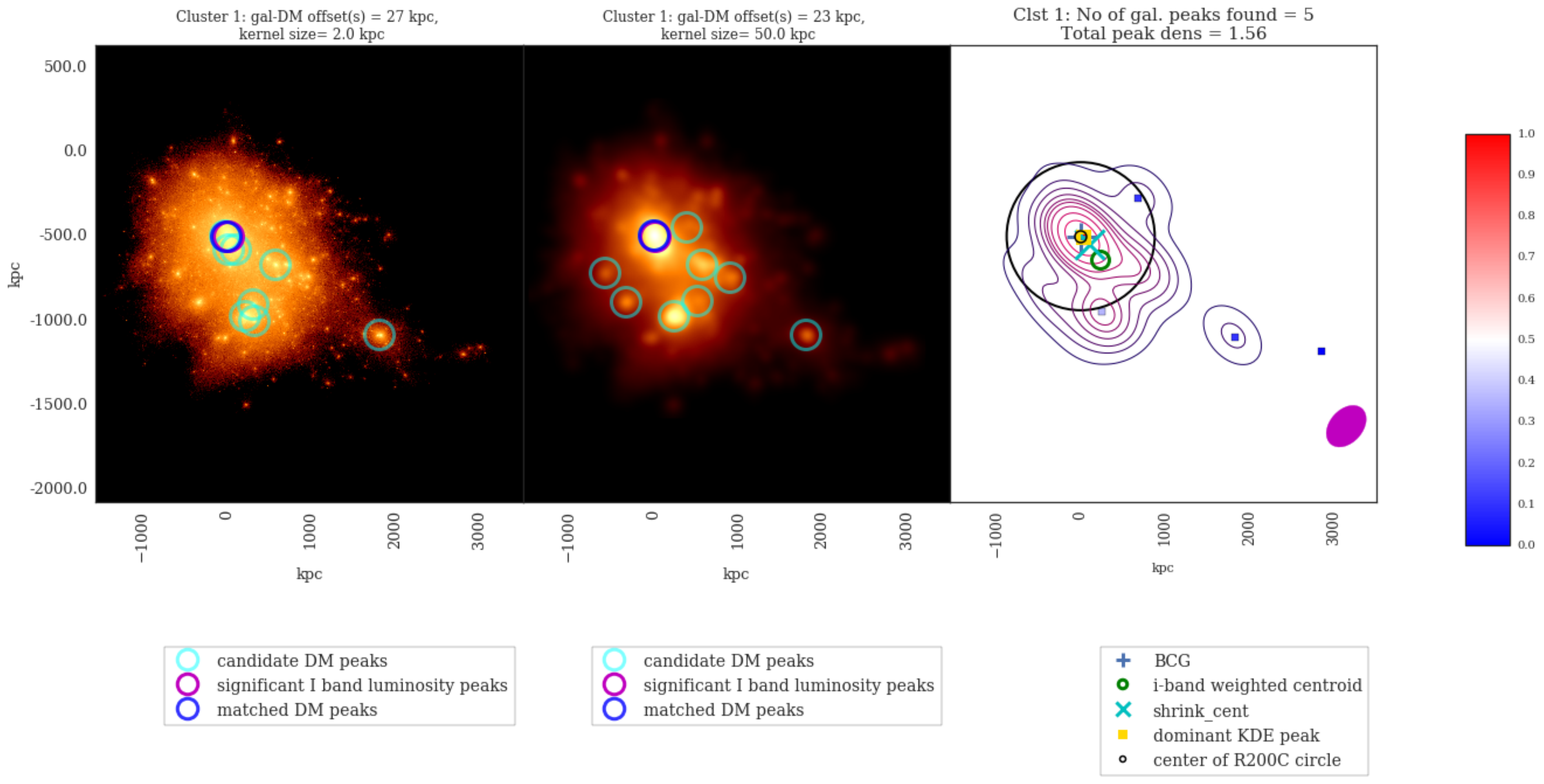}}{}
	\includegraphics[width=0.78\linewidth]{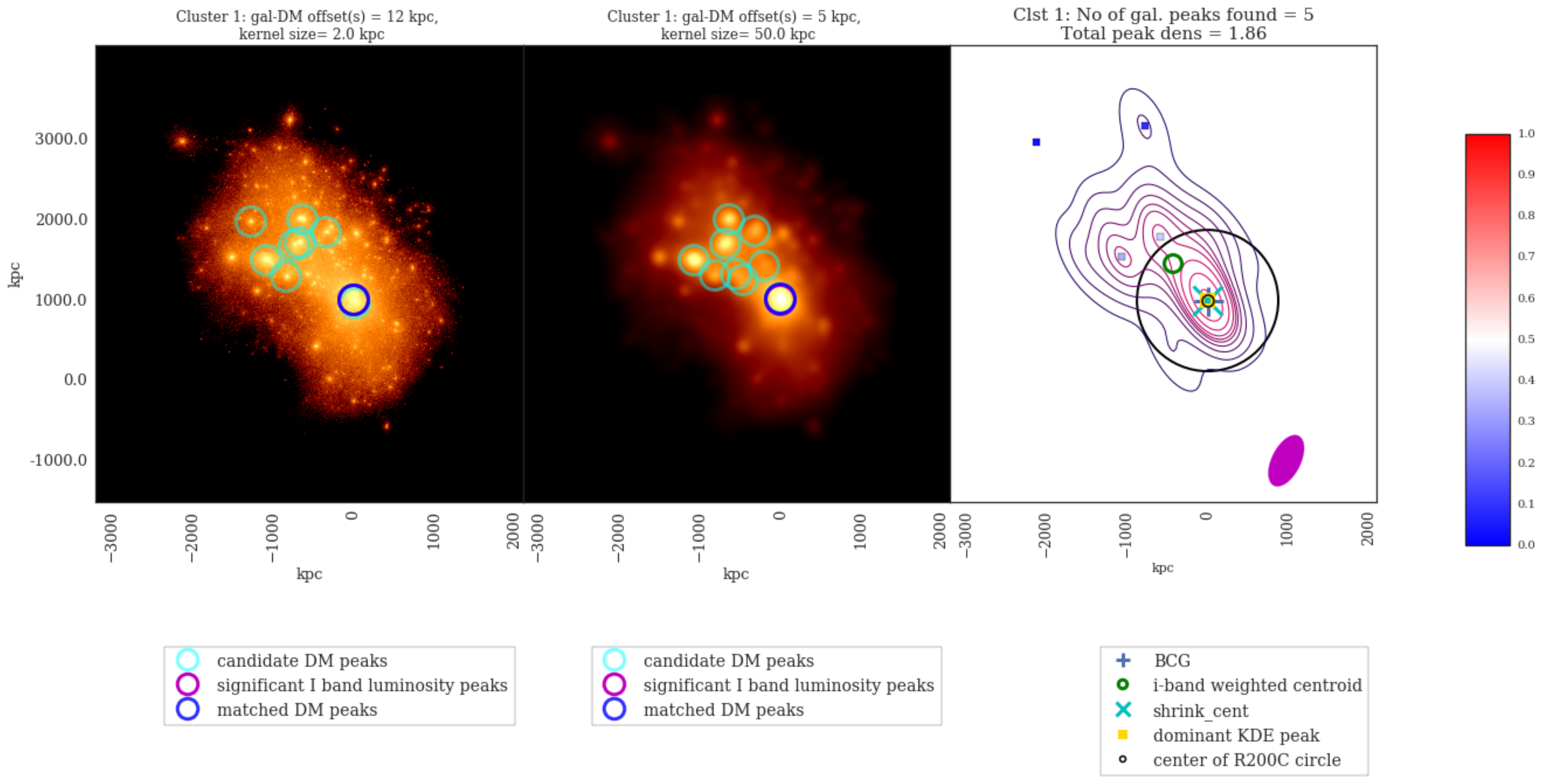}
	\includegraphics[width=0.78\linewidth]{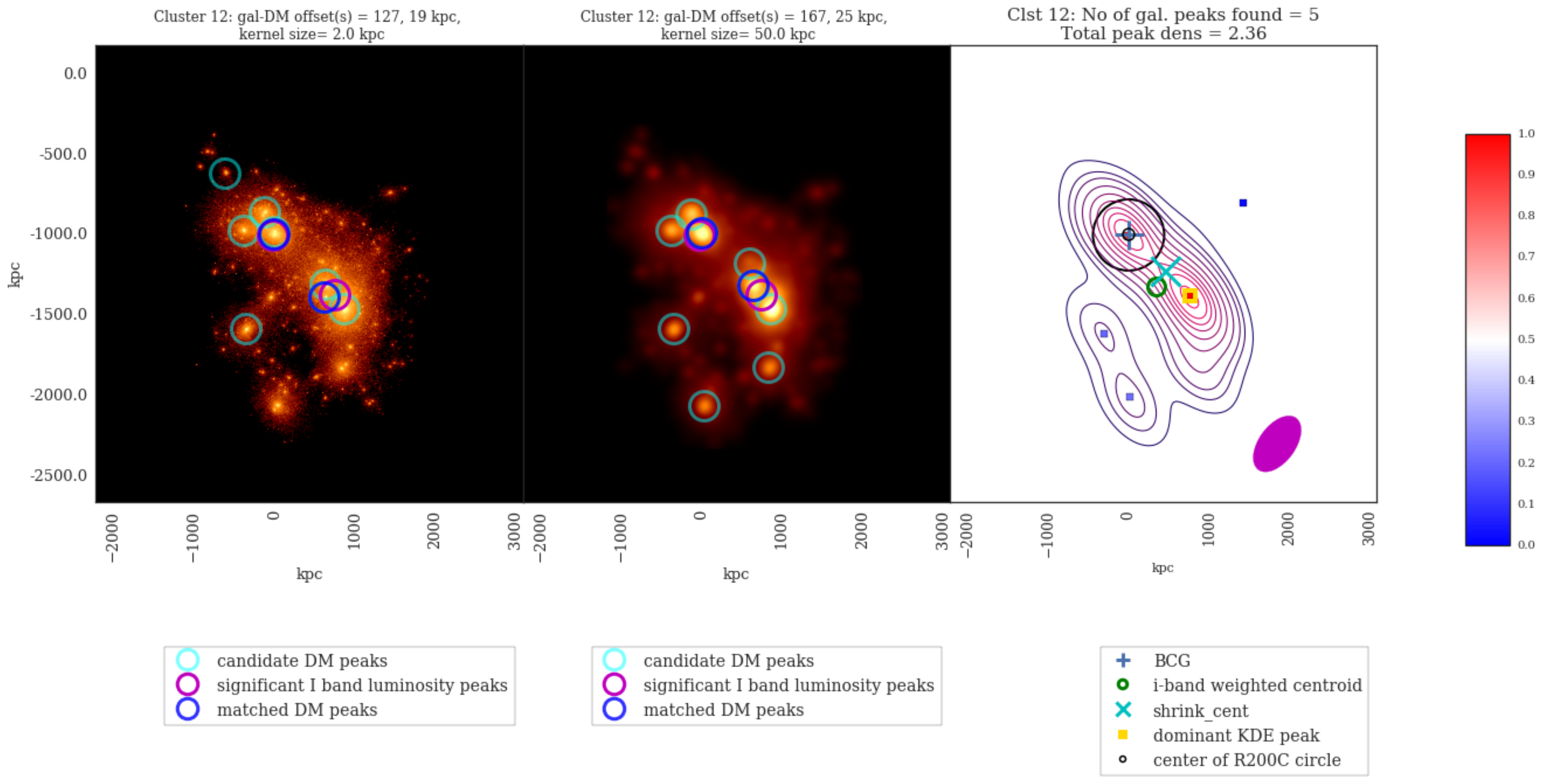}
	\caption{Visualization of Illustris clusters, each row representing a given projection
		of a given cluster (top two rows show two different projections of the same cluster). Left and mid columns: Projected density distributions of DM	
		particle data are shown in the yellow-orange-red color scheme, with the densest regions in yellow, for two different choices of the smoothing kernel (2 and 50 kpc, respectively).
	         The identified DM density peaks (local maxima) are indicated by colored circles.
		Right column: Projected galaxy kernel density estimates (KDE) of 
		the $i$-band luminosity map for the member galaxies of the same clusters. Each colored contour denotes a 10\% drop 
		in density starting from the highest level in red. Each magenta ellipse on the
		bottom right corner of the panels shows the Gaussian kernel matrix 
		$H$ from eq. (\ref{eq:cross_validated_bandwidth}). 
		The big black circle is centered on the most bound particle identified by {
		\sc SUBFIND } and the radius of the circle indicates 
		R$_{\rm 200C}$. The KDE luminosity peaks (square markers) are colored
		by their relative density to the densest peak (normalized KDE density -- see color bar on the right and text and Section \ref{subsubsec:KDE} for details). See Appendix \ref{app:visualization} for more examples.
		\label{fig:select_peak_visualization}
	}
\end{center}
\end{figure*}

\subsection{Finding the DM summary statistics}
\label{subsec:DM_map}
The most well established method for inferring the projected dark matter spatial 
distribution from observations is through gravitational lensing.
It works by detecting subtle image distortions of background galaxies due to
the foreground dark matter. Observationally, the resolution of the inferred map therefore 
depends on the properties of the source galaxies that are being lensed, 
such as the projected number density, the intrinsic ellipticities and morphology etc.\\
Here, in order to estimate the DM spatial distribution of Illustris clusters and mimic the lensing-based DM peak measurements, we directly use the DM particles from the simulation. We first make a histogram of the DM particle positions with 2 kpc
$\times$ 2 kpc bin size, which is slightly larger than the DM softening length of 1.4 kpc. 
Then, we use a (50, 50) kpc 2D radial Gaussian kernel to smooth the DM histogram made from the Illustris DM
particle data.
The choice of 50 kpc is dictated by observational methods.
\cite{Hoag2016} has performed a simulation for inferring the optimal bandwidth
for a Gaussian smoothing kernel for the cluster MACSJ0416: in the strong lensing regime, they found it to be 11 arcseconds.
A kernel bandwidth (this is the standard deviation) 
of 11 arcseconds translates to an angular diameter distance of 50 
kpc assuming a cosmological redshift of $z \approx 0.3$. 
In order to match the resolution of lensing data,
we also therefore employ a smoothing kernel of a similar physical size of 50 kpc.  However, it should be noted that MACSJ0416 is one of the best-studied clusters, 
with many strong lensing constraints, and that clusters without strong lensing constraints will be mapped with 
lower resolution. We purposely choose our resolution to study the best-case scenario, that observers will be able in turn to degrade according to the quality of their data.

As it can be seen from Fig. \ref{fig:select_peak_visualization} (left two columns), there are non-negligible differences between the smoothed and unsmoothed DM
maps. The unsmoothed histograms tend to show many more local maxima around the major
density peaks (i.e. show high variance), with millions of DM particles for each cluster, densely packed in the region of
interest. \\

Our working assumption is that the smoothed DM histograms of our simulated clusters 
are good proxies for the convergence maps from high-resolution lensing analyses in the limit of perfect foreground/background rejection. We again employ standard algorithms to find the local maxima of the DM smoothed density distribution, which we call DM peaks.

\subsection{Finding the offsets} \label{subsec:offsets}
Finding the offsets between the peaks of the DM maps and the summary statistics of the 
galaxy population is not a well-defined process; this is even more problematic in the presence of multiple peaks.

We refer the reader to Fig. \ref{fig:select_peak_visualization} for a comparison
between the DM maps (e.g. left and mid columns, for different smoothing kernel sizes), and 
the KDE of the luminosity maps (right column).
Each row depicts maps for a given Illustris cluster in a given projection.
There can be multiple peaks in each map due to substructures. 
There are also many more DM peaks (indicated by circles on the left two columns) 
than luminosity peaks (indicated by the squared markers on the right column).
This is because there are many more dark subhaloes than galaxies for each cluster and 
the resolution of the DM data is much higher. 
Furthermore, due to projection, not all peaks correspond to actual substructures.\\

Clearly, different peak identification criteria and different procedures to match DM and galaxy summary statistics can lead to substantially different offset values. 
Below, we outline the procedure we adopt to match the DM peaks (obtained as described in Section \ref{subsec:DM_map}) to the peaks of the luminosity maps determined for every cluster projection as described in Section \ref{subsubsec:KDE}: we will then use the {\it matched DM peaks} to compute offsets with any other galaxy summary statistics defined in Section \ref{subsec:methods:galaxy}.

\subsubsection{Matching DM and luminosity peaks}

We write our peak matching algorithm to mimic what humans would do to find the ``nearest important match'' between the DM and galaxies peaks.\\

For the galaxy peaks, the KDE method provides a natural ranking procedure. Namely, as already anticipated in Section \ref{subsubsec:nu}, the importance of the galaxy peaks in a cluster is determined by the density 
estimate at each luminosity peak location, normalized to the value of the brightest one in that cluster. In Fig. \ref{fig:select_peak_visualization}, rightmost column, this is indicated by the color of the square markers, according to the color bar on the right\footnote{There are some spurious substructures as indicated by 
the deep blue squared markers in the right hand panels of Fig. \ref{fig:select_peak_visualization}. They represent peaks created by a small number of galaxies that
are located far away from the main concentration of mass.}. We call {\it significant} those luminosity peaks with normalized density exceeding 0.2, and indicated with magenta circles in the left and mid columns of Fig. \ref{fig:select_peak_visualization}. \\

For the DM peaks, the method of Section \ref{subsec:DM_map} returns multiple peak locations, 
as indicated by the cyan or blue circles in left and mid columns of Fig. \ref{fig:select_peak_visualization}. Also DM peaks can be ranked according to the density at their locations.\\

For each significant luminosity peak, we look for the closest DM peak by using a data structure called a k-dimensional tree (KD-Tree; in
our case, k = 2). A tree stores the 2D locations of the DM peaks in a sorted
manner and can speed up the identification of the closest DM peak from the
location of the luminosity peaks.
We do not compute the distances
between all possible pairs of DM and significant galaxy peaks. 
In practice, for those clusters with multiple significant luminosity peaks, we limit the number of DM peaks to:
\begin{equation}
	N_{\rm DM} = \begin{cases}		
		3 \times (N_{\rm lum} + 1) & {\rm if~} N_{\rm lum} < 3 \\
	3 \times N_{\rm lum}  & {\rm if~} N_{\rm lum} \geq 3.
	\end{cases}
	\label{eq:peak_threshold}
\end{equation}
where $N_{\rm lum}$ is the number of significant luminosity peaks and $N_{\rm DM}$
is therefore the number of DM peaks included for the construction of the KD-tree. The DM peak that minimizes the 2D distance from each significant luminosity peak is then identified as the {\it matched DM peak}.
Notice that, when there are several dense galaxy peaks located far away from one another, 
the top few densest DM peaks (subhaloes) can locate around the same galaxy
peak (see the cyan circles in Fig. \ref{fig:select_peak_visualization} for the
third row), i.e. there is no one-to-one matching between the luminosity of 
galaxies and the density of detected DM peaks. Hence the procedure defined with eq. \ref{eq:peak_threshold} is effective.
We have checked from the visual inspection of figures analoguos to
Fig. \ref{fig:select_peak_visualization} that using eq. \ref{eq:peak_threshold} works well to match the 
appropriate peaks, i.e. the blue circles that indicate the matched DM peak 
in the left and middle columns 
often overlap the magenta circle representing the significant luminosity peaks.

\subsubsection{Measuring the galaxy-DM offsets}
To prevent peak matching from affecting the offset results, for each cluster projection we quote and use in what follows only one offset measure per galaxy summary statistics. Namely, among the matched DM peaks identified above, we keep only the DM peak matched to the brightest i.e. densest luminosity peak. 

We therefore measure the 2D distances between such matched DM peak and the following spatial estimates, with corresponding offset labels in parentheses:

\begin{itemize}
	\item the most (gravitationally) bound particle
	\item the BCG ($\Delta s_{\rm BCG}$);
        \item the luminosity weighted centroid ($\Delta s_{\rm cent}'$)
	\item the shrinking aperture centroid ($\Delta s_{\rm shrink}'$) 
	\item the brightest KDE luminosity peak ($\Delta s_{\rm KDE}'$)
	\item the closest KDE number density peak ($\Delta s_{\rm num.dens}$)\footnote{As there can be more than one number density peak from the corresponding KDE
map, we also use a KD-tree to match the closest number density peak to the 
identified DM peak.}.
\end{itemize}


The prime symbol $'$ in the offset labels is a reminder to that $i$-band luminosity weights have been adopted in the measure of the galaxy summary statistics.
From all selected clusters and projections, in Section \ref{sec:results}, we will report percentiles of the offset distributions as well as other statistics of interest, such as the biweight 
location (analogous to the median) and the midvariance (robust standard deviation estimate) to minimize the effects of outliers \citep{Beers90}. 
For instance, the 95\% interval is computed as the narrowest interval that encompasses
95\% of total density (2.5\% of density mass at each end of the tail is
excluded). In case of degeneracy, the interval is also required to cover the 
central location estimate for the distribution.
We compute the robust statistics using the 
implementation of {\bf \texttt{astropy.stats.biweight\_location}}
and {\bf \texttt{astropy.stats.biweight\_midvariance}}
from \cite{astropy} as part of {\sc{astropy}}.

\subsection{2D vs 1D Offsets}
\label{subsec:1d}

In the following sections of this \ifthenelse{\boolean{thesis}}{chapter}{paper}, 
we use $\Delta {\bf s}$  to represent the
two-dimensional offsets, $|\Delta {\bf s}|$ for the magnitude of the offset as calculated
according to the Euclidean distance, and $\Delta x$ or $\Delta y$ to denote the
one-dimensional offset along one of the spatial dimensions.

The most faithful representation of the offsets without any information loss
is the following 2D form:
\begin{equation}
	\Delta {\bf s} = ({\bf x}_{\rm gal} - {\bf x}_{\rm DM}, 
	{\bf y}_{\rm gal} - {\bf y}_{\rm DM} ).
	\label{eq:2D_offsets}
\end{equation}
Its probability distribution function peaks at (0, 0) when there is no real offset. By taking the magnitude of $\Delta {\bf s}$, i.e.:
\begin{equation}
	|\Delta {\bf s}| = \sqrt{({\bf x}_{\rm gal} - {\bf x}_{\rm DM})^2 + 
	({\bf y}_{\rm gal} - {\bf y}_{\rm DM})^2},
	\label{eq:magnitude_offsets}
\end{equation}
the resulting 1D distribution of $|\Delta {\bf s}|$, 
whose support being [0, $\infty$),
will not peak at zero even if the original
distribution of $\Delta {\bf s}$ peaks at (0, 0). It is indeed difficult to interpret the magnitude values of $|\Delta
s|$ given $\Delta {\bf s}$. For example,  the 1D $|\Delta s|$ values corresponding to a 2D standard Gaussian centered in zero with standard deviation $\sigma$,
\begin{equation}
	\left(\begin{array}{c}
			\bf{x}\\
			\bf{y}
		\end{array}
	\right) \sim \mathcal{N}\left(
	\left(
		\begin{array}{c}
			0 \\
			0
		\end{array}
	\right),
	\left(\begin{array}{cc}
		\sigma^2, 0 \\
		0, \sigma^2
	 \end{array}
	\right)
\right),
\end{equation}
follow a Rayleigh distribution:
\begin{equation}
	f(\Delta s | \sigma) = \Delta s /  \sigma^2 \exp(-\Delta s^2 / 2 \sigma^2),
	\label{eq:rayleigh_distro}
\end{equation}
with peak $|\Delta s| = \sigma$.
However, the dependency of $|\Delta s|$ on the parameters of
the 2D distribution is much more complicated when the 2D distribution does not approximate a Gaussian 
or when there is more than one peak in the 2D space (see Table \ref{tab:offset_distributions} for a comparison between 1D and 2D offsets from Illustris clusters).

In what follows, we will therefore report results also for the 1D distributions of offsets along a particular spatial axis, 
$\Delta {\bf x}$ and $\Delta {\bf y}$, which, having 
each a support of $\mathbb{R}$, will not exhibit a discontinuity at zero, and are hence to be favored to the 2D offsets.
Since we have enough cluster projections for there to be
rotational symmetry for the distribution of $(\Delta {\bf x}, \Delta {\bf y})$, it does not
matter if we pick $\Delta {\bf x}$ or $\Delta {\bf y}$ for the 1D representation.

\begin{figure*}
	\centering
	\includegraphics[width=.85\linewidth]{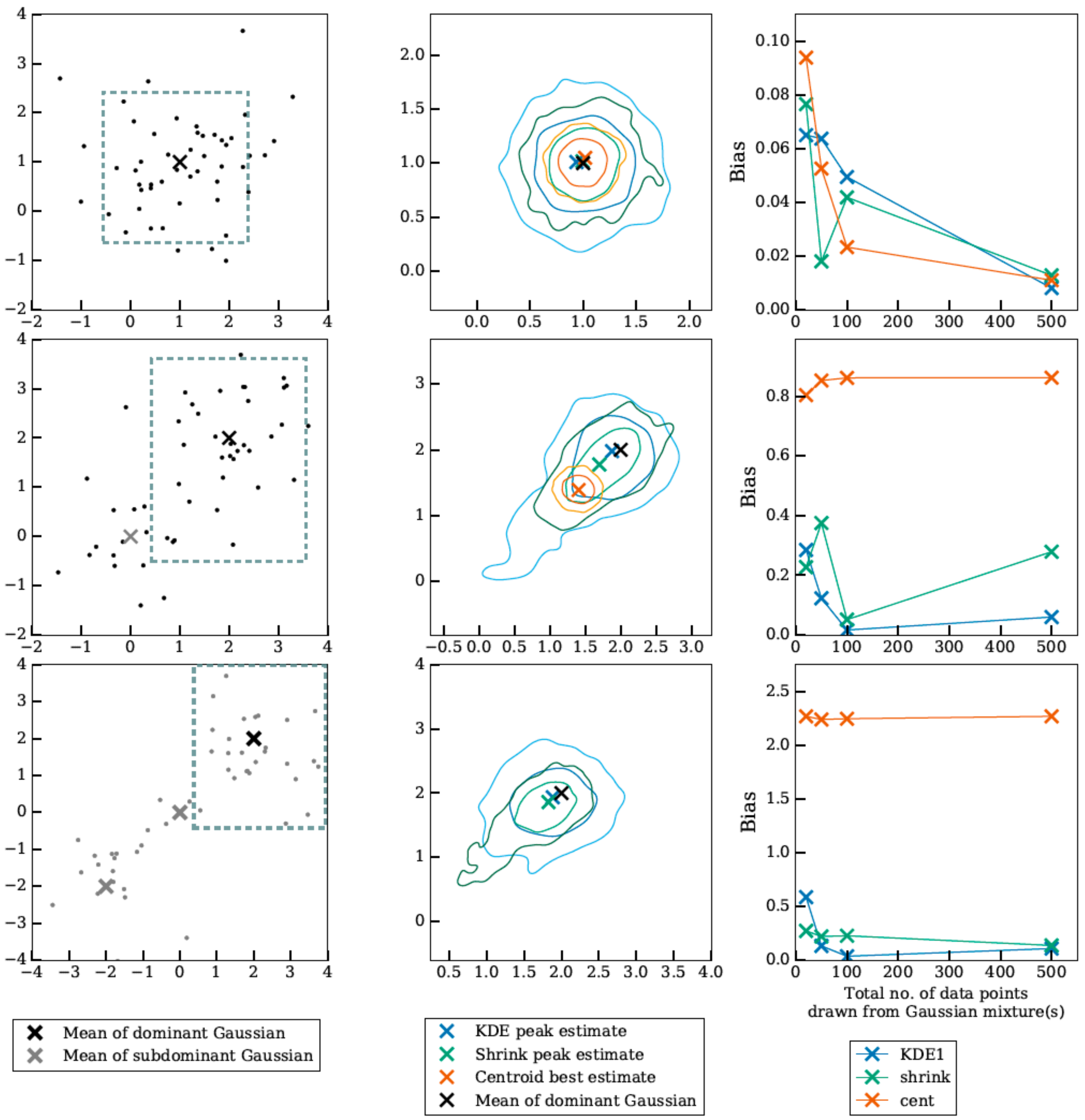}
	\caption{Idealized comparisons of galaxy summary statistics. Here we compare the performances of the shrinking aperture
		centroid (shrink peak), dominant peak estimate from the KDE map (KDE1), and the centroid (cent), by
		drawing data points (i.e. 20, 50, 100, 500) from a known number of 
		Gaussian mixtures. 
		From top to bottom, we use data drawn from a single Gaussian mixture (top row), two 
		 Gaussian mixtures with weight ratio = 7:3 (mid row), and three Gaussian
	mixtures with weight ratio = 55:35:10 (bottom row).
	The left column shows a realization of 50 data points drawn from the Gaussian
	mixture(s). Zoomed-in views of the data, shown by the dash outlined boxes, are given in the middle column.	
	Due to the statistical nature of this exercise, we sample the data and
	perform the analyses 1000 times to
	create the 68\% and 95\% Monte Carlo confidence contours of the estimates in the
	middle column. The rightmost column shows how the population bias varies as a
	function of the number of drawn data points from the Gaussian mixtures, where here the bias is defined as the 2D distance between each method's summary statistics and the true gaussian centres.
	The middle and rightmost column demonstrate that, in the cases of multiple gaussians/peaks and for enough data points, the cross-validated KDE peak estimate is the most accurate method (for numbers of data points $\gtrsim 50$).  
		\label{fig:toy_data_mixtures}}
\end{figure*}

\section{Validation of the Methods} 
\label{sec:validation}

\subsection{Idealized comparisons of the galaxy summary statistics}
We examine the properties and functioning of the galaxy summary statistics by testing them on synthetic data drawn
from Gaussian mixtures with known mean and variance. 
Results are shown in Fig.
\ref{fig:toy_data_mixtures} for KDE peak estimate, shrinking aperture and centroid methods. 
The main factors that affect the performance of 
the methods are 1) the actual density profiles, 2) the location(s) of subdominant mixtures, and 3) the number of data points that we draw.
Due to statistical fluctuations, it would not be enough to
compare the performance of the galaxy summary statistics by applying each method for just one realization of the
data. We hence provide the 68\% and the 95\% confidence regions by applying
each method for 1000 realizations from the Gaussian mixtures.
We compute the population location from the 1000 realizations for each method 
and indicate it as a cross on the middle column of Fig. \ref{fig:toy_data_mixtures}.
In general, the peak identified from the KDE density is closer to the 
peak of the dominant mixture (more accurate) than 
the weighted centroid method and the shrinking aperture method.
For example, in the bottom middle panel of Fig. \ref{fig:toy_data_mixtures}, 
the green contours that represent the confidence region for the shrinking aperture peak are
biased due to the presence of the multiple gaussians, representing multiple cluster substructures, whereas the confidence region for the centroid 
is so biased that it falls outside the field of view of that panel.
In the right panel of Fig. \ref{fig:toy_data_mixtures}, 
we present how the population bias of each method decreases as the
number of data points increases. We conclude that, with enough data points ($> 50$), 
for the data generated with more than one mixture, 
the KDE peak consistently shows less population bias than the shrinking aperture method. 
The performance of the shrinking aperture method fluctuates and is unstable when
the number of data points is increased.

\subsection{Offset between DM peak and most
bound particle}
\label{subsec:offset_from_most_bound_particle}
Before discussing the galaxy-DM offsets of Illustris clusters, as a sanity check we comment on the distance between the simulation's most gravitationally bound particle and the various summary statistics defined in Section \ref{sec:methods}.
In each cluster and projection, there is no significant offset between the matched DM peak and 
the most gravitationally bound particle (hereafter most bound particle, which in the simulated clusters could be of any matter type: a stellar or DM particles, a gas cell).
The median DM-most bound particle offset is (0, 0) kpc, with 75-th percentiles of ($\pm2,\pm2$) kpc. 
Most of the other offset values occur below ($\pm 9, \pm 9$) kpc. Large offsets
are only seen for clusters with $\nu > 1.2$: the densest DM peak in 3D where
the most bound particle is located does not necessarily correspond to the densest projected peak in 2D in the presence of 
significant DM substructures.   

\subsection{Offset between galaxy summary statistics and most
bound particle}
As another sanity check, we compute the offsets between different galaxy summary
statistics and the most bound particle in each projection. Results are summarized in Table
\ref{tab:most_bound_particle_offset_distributions} for the different 
percentiles and robust estimates of the distributions.
The various galaxy methods are ranked as follows, from the closest to the farthest from the most bound particle:
1) the BCG position; 2) the densest KDE luminosity peak; 3) the shrinking aperture centroid ($i$-band weighted); 4) the closest KDE number density peak (unweighted); and 5) the luminosity weighted centroid.

In fact, most of the BCG offsets are very small except for the two clusters with ID 13
and 33. Both clusters have  values of $\nu > 1.5 $ over each and every projection, and visual inspection of the projected luminosity and DM maps confirms that
both clusters have significant substructures. It is therefore possible for the
most bound particle to have a similar gravitational potential level as another 
substructure where the BCG is located. 
In general, the offset distributions between the galaxy summary statistics and
the most bound particle have approximately the same level of variance but more
extreme outliers (at the 99\% level) than the
offset distribution between the DM peaks and the corresponding galaxy summary
statistics (next Section).

\ifthenelse{\boolean{thesis}}{\begin{landscape}}{}
\begin{figure}
\begin{center}
	\includegraphics[width=0.85\linewidth]{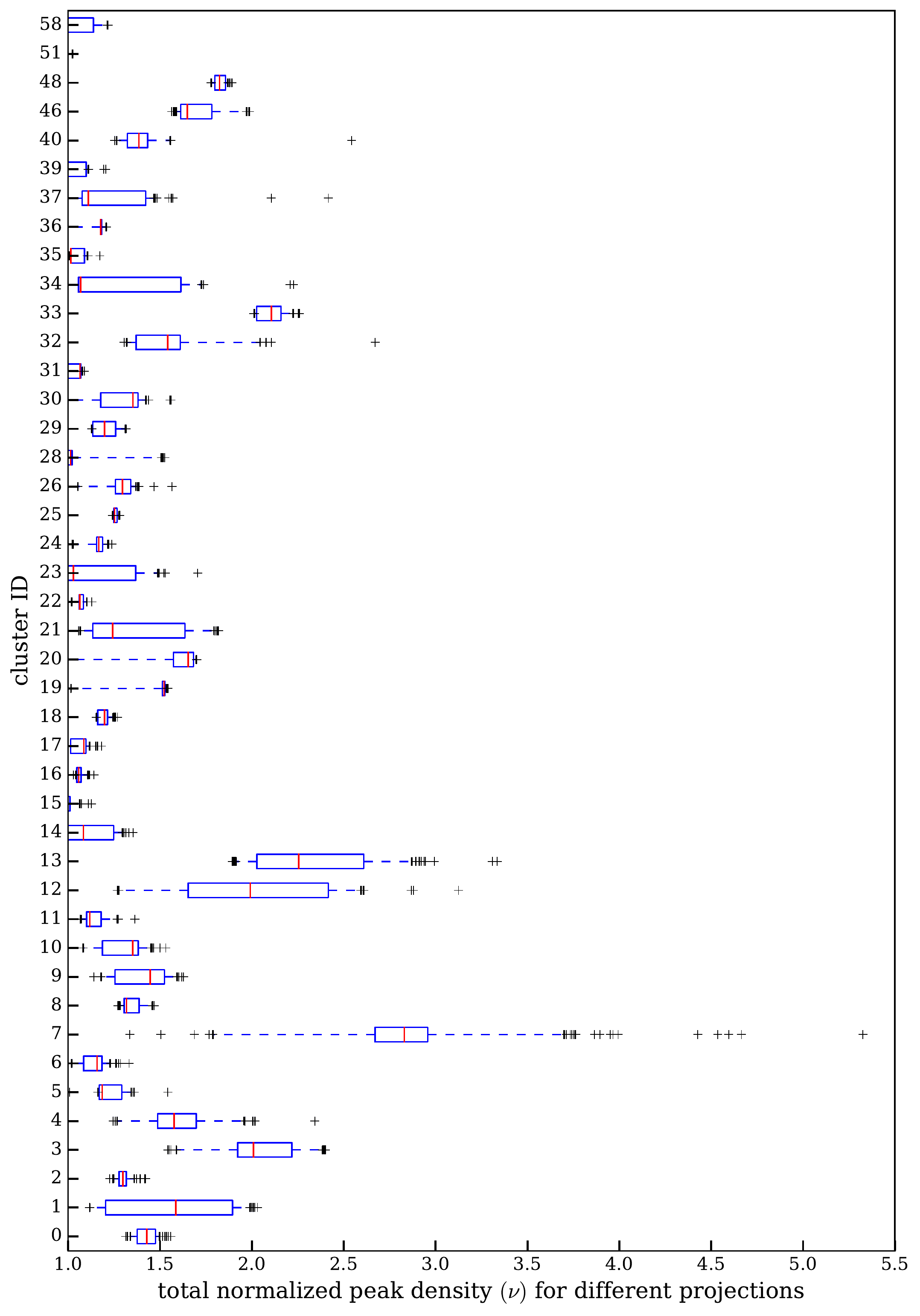}
	\caption{Distribution of the total normalized peak density ($\nu$)
		for each cluster selected from Illustris. $\nu$ is obtained from the KDE maps and is a good proxy for relaxedness. Clusters with only one dominant peak have $\nu = 1.0$, and
		values bigger than 1 indicate that the total density is contributed by more than one significant peak. The red line of each box shows the median of the projections,
		the box encompasses the 25\% and 75\% percentile of the distribution while
		the whiskers mark the 5\% and the 95\% percentile. The other black crosses
		are data points with extreme values beyond the 5\% and 95\% percentile.
		\label{fig:nu_distribution} 
	}
\end{center}	
\end{figure}
\ifthenelse{\boolean{thesis}}{\end{landscape}}{}

\begin{figure*}
	\centering
	\includegraphics[width=0.65\linewidth]{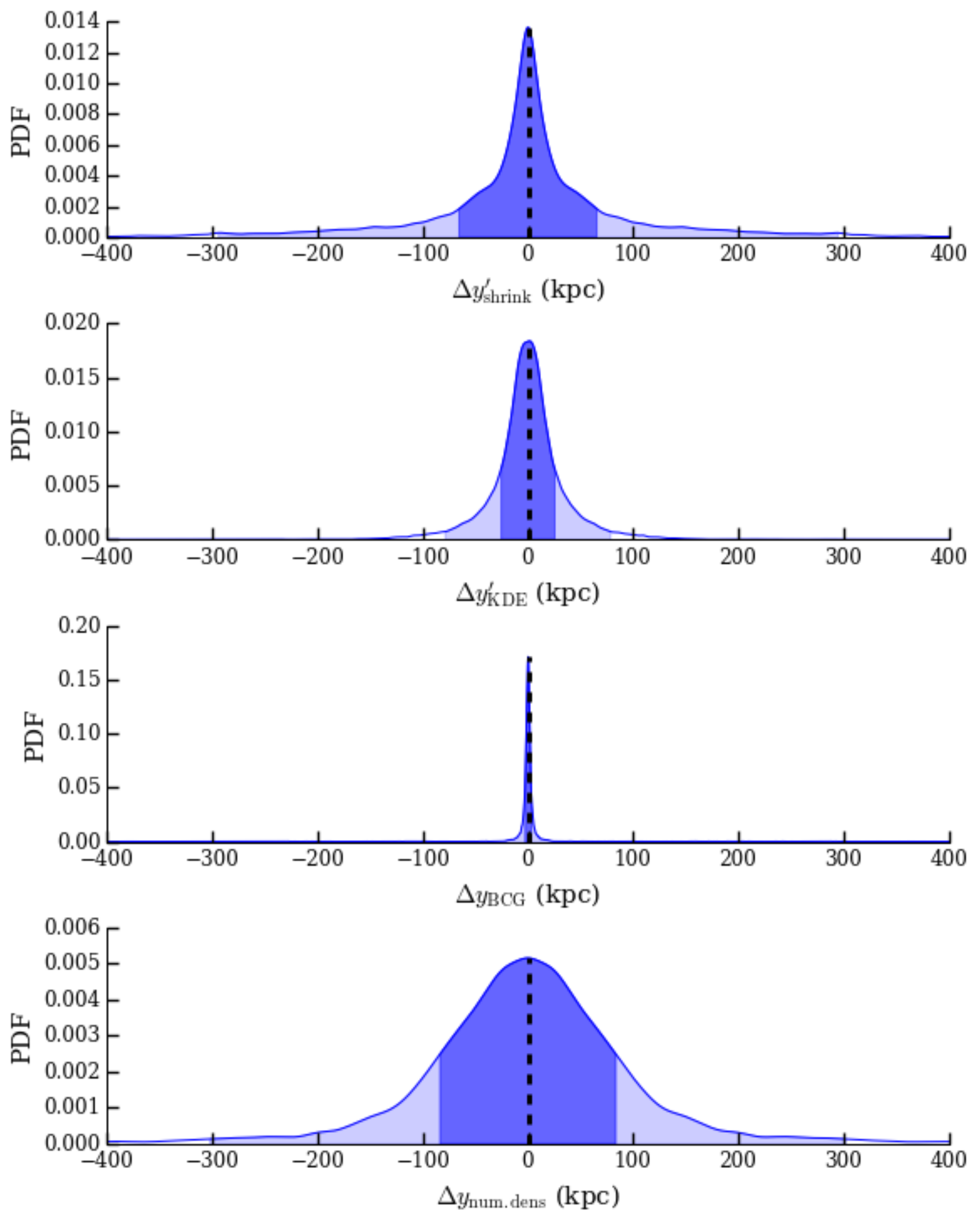}
	\caption{Galaxy-DM offset distributions from 43 Illustris, $\Lambda$CDM  clusters analyzed in 33 024 projections: here we show the smoothed distributions of the 1D offsets along the y-axis, for the different methods usually adopted to determine the galaxy summary statistics (see Section \ref{sec:methods} for details -- smoothing bandwidth determined by Scott's rule \citep{Scott2010} for visualization). A summary of the statistics of each distribution is available in Table \ref{tab:p_val_table}. We omit the case where the galaxy statistics is determined via the luminosity-weighted centroid.
	For estimates where several peaks of galaxy data are 
	possible, only the densest peak is matched to the DM peak for measuring
	the offsets. The dark blue area indicates the 68\% density interval
	while the light blue area shows the 95\% density interval. 
		\label{fig:offset_distributions}}
\end{figure*}
\section{Results} 
\label{sec:results}

\begin{table*}
	\begin{center}
		\caption{Galaxy-DM offset distributions from 43 Illustris, $\Lambda$CDM clusters analyzed in 33 024 projections. Robust estimates and distribution properties are given for the 1D offsets along the y-axis (as in Figure \ref{fig:offset_distributions}), for the different methods usually adopted to determine the galaxy summary statistics (see Section \ref{sec:methods} for details). The sample is divided in relaxed ($\nu < 1.2$) and unrelaxed ($1.2 < \nu < 2.2$) clusters according to the definitions and findings of Sections \ref{subsubsec:nu} and \ref{subsec:nu_results}. Sections \ref{subsec:hypothesis_test} and \ref{subsec:1d} explain how $\Delta y$ is different from the offset magnitude which has a discontinuity at zero: results for the 2D offset distributions are summarized in Table \ref{tab:offset_distributions}.
		\label{tab:p_val_table}
	}
	\begin{tabular}{llccccccc}
\toprule
sample & offset (kpc) &  location &  lower 68\% &  lower 95\% &  lower 99\% &  upper 68\% &  upper 95\% &  upper 99\% \\
\midrule
all $\nu$ & $\Delta y_{\rm BCG}$               &         0 &          -3 &         -22 &        -496 &           3 &         456 &        1449 \\
all $\nu$ & $\Delta y_{\rm KDE}'$              &         0 &         -25 &         -79 &        -127 &          25 &          79 &         126 \\
all $\nu$ & $\Delta y_{\rm centroid}$     &  0 &         -181 &-563 &        -826 &          181&         563&         826\\
all $\nu$ & $\Delta y_{\rm num. dens}$         &        0 &         -84 &        -303 &        -693 &          84 &         302 &         691 \\
all $\nu$ & $\Delta y_{\rm shrink}'$           &        0 &         -65 &        -295 &        -652 &          65 &         295 &         655 \\
\midrule
$\nu < 1.2$ & $\Delta y_{\rm BCG}$             &        0 &          -3 &         -10 &         -19 &           2 &           9 &          19 \\
$\nu < 1.2$ & $\Delta y_{\rm KDE}'$            &         0 &         -18 &         -48 &         -82 &          18 &          48 &          83 \\
$\nu < 1.2$ & $\Delta y_{\rm centroid}'$        &        0 &        -108 &        -255 &        -395 &         108 &         254 &         394 \\
$\nu < 1.2$ & $\Delta y_{\rm num. dens}$       &        0 &         -73 &        -195 &        -303 &          73 &         195 &         302 \\
$\nu < 1.2$ & $\Delta y_{\rm shrink}'$         &         0 &         -51 &        -187 &        -285 &          51 &         187 &         285 \\
\midrule
$1.2 < \nu < 2.2$ & $\Delta y_{\rm BCG}$       &         0 &          -3 &        -160 &        -684 &           4 &         807 &        1570 \\
$1.2 < \nu < 2.2$ & $\Delta y_{\rm KDE}'$      &        0 &         -32 &         -89 &        -125 &          32 &          89 &         124 \\
$1.2 < \nu < 2.2$ & $\Delta y_{\rm centroid}'$  &         0 &        -262 &        -663 &        -905 &         262 &         663 &         904 \\
$1.2 < \nu < 2.2$ & $\Delta y_{\rm num. dens}$ &         0 &         -87 &        -299 &        -739 &          87 &         298 &         738 \\
$1.2 < \nu < 2.2$ & $\Delta y_{\rm shrink}'$   &        0 &         -85 &        -386 &        -777 &          85 &         386 &         779 \\
\bottomrule
\end{tabular}

\end{center}
\end{table*}

\subsection{The dynamical states (relaxedness) of the clusters}
\label{subsec:nu_results}
Out of the 43 clusters $ \times$ 768 projections / cluster  = 33 024 projections, $\sim 45\%$ of them
have one dominant luminosity peak and negligible substructures, with the total peak density of the
projection being $\nu \leq 1.2$. 
Another $\sim 50\%$ have more than one dominant luminosity
peak with $1.2 < \nu < 2.2$, and about 5\% have an abundance of substructure with $\nu>2.2$. This is shown in Fig.
\ref{fig:nu_distribution} adopting the definition of total normalized peak density $\nu$ given in Section \ref{subsubsec:nu}.
Visually, the spread of the $\nu$ distribution is indicated by the horizontal 
length of the blue box. 
The median $\nu$ per cluster is indicated by the red central vertical line
inside each box.
Only 7 clusters (with ID = 15, 16, 17, 22, 31, 35, 51) out of 43 clusters have $\nu
\lesssim 1.2$ for most of the projections.
Clusters with median values of $\nu > 2.2$ usually have multiple subclusters.
The cluster with ID = 7, for instance, is made up of around 4 sub clusters that span
several Mpc.  

We have checked the connection between $\nu$ and the definitions of relaxedness typically adopted with simulation data: there is a strong correlation of $\sim 0.8$ between the value of $\nu$ averaged across projections and
each of the two unrelaxedness quantities defined in Section \ref{subsubsec:relaxedness}. This result confirms that $\nu$ is a good indicator of the dynamical state of the cluster and hence a useful a proxy which can be applied to observed data.
In Table \ref{tab:cluster_prop} in Appendix, the values of our measures of relaxedness are given for each selected Illustris cluster. In Figure \ref{fig:mass_abundance_distribution}, the mass distributions of the Illustris selected clusters are given for different relaxedness selections: the high mass end of the selected sample ( $ \gtrsim 1.5 \times 10^{14}$ M$_\odot$ ) is dominated by unrelaxed clusters with multiple subcomponents.

\subsection{Galaxy-DM Offsets in Illustris}
\label{subsec:galaxyDMoffset}

\subsubsection{The two-dimensional offset distributions $\Delta s$}

The 2D offset distributions of Illustris clusters are recorded in detail in Table
\ref{tab:offset_distributions}, denoted by $\Delta s$.
For most methods, the 2D distributions peak around zero ($\lesssim 4$ kpc), across projections.
The method that gives the tightest offset is the BCG. 
The 2D offset $\Delta s_{\rm BCG}$ has most of its density located near zero
($\pm 3$ kpc) but contains outliers. Having outliers is possible 
as the DM peak is chosen as the closest DM peak to match the
brightest luminosity peak in a particular projection. See for example the bottom right panel of Fig. \ref{fig:select_peak_visualization}: there the BCG 
coincides with the most bound particle. However, the luminosity peak of the
cluster is located at the other mass substructure.
When there are distinctly separated subclusters of similar masses, 
the brightest projected luminosity peak may shift from one subcluster to another across different projections,
while the BCG identification is unchanged between projections.

On average, the 2D offset distributions are consistent across projections, namely there is a good rotational symmetry.
Possible sources of offset asymmetry arise with clusters with unusual configurations, e.g. those with more distinct, spatially separated subcomponents.
In those cases, the offset variance may differ significantly across projections, and outliers may be more extreme than in more relaxed clusters.

\subsubsection{The one-dimensional offset distributions $\Delta y$}

As discussed in Section \ref{subsec:1d}, the 1D offsets along a given projected axis ($\Delta y$ or  $\Delta x$) are better defined and robust quantities than the 2D analogs. We therefore report in detail our results from Illustris in terms of $\Delta y$: the main results of this paper are given in Table \ref{tab:p_val_table} and Figure \ref{fig:offset_distributions}, where the offset distributions are obtained across 33 024 projections of 43 Illustris clusters.

From the different rows of Fig. \ref{fig:offset_distributions}, we can see 
that the variance from each offset method is very different. First of all, it is therefore unreasonable to compare offsets that are generated 
by different methods of peak inference across studies.
For the full sample, the offsets computed with the BCG method have the smallest variance, followed by the luminosity weighted KDE method. The 68-th percentiles of $\Delta y_{\rm BCG}$ and $\Delta y_{\rm KDE}'$ are at $\pm 3$ kpc and $\pm 25$ kpc, respectively. Using shrinking aperture to estimate
the peak location from the luminosity map increases the 68-th percentiles by at least a factor of two, $\pm 65$ kpc.
The peak estimate from the number density map has even larger variance ($\pm 84$ kpc) and the centroid method is in practice unusable.\\

The aforementioned bias from substructure can be seen when we compare the
offset estimates between the relatively relaxed sample  ($\nu < 1.2$) and the
unrelaxed samples ($1.2 < \nu < 2.2$). 
For the relaxed sample, the variance of the offsets inferred from different
methods still shows significant discrepancies. 
The variances of the offsets measured with the shrinking aperture method, 
the number density map, and the weighted centroid are still at least a factor of 1.5
larger than those from using the luminosity-weighted KDE. 
In particular, the 68\% percentile of the centroid method is $\pm 108$ kpc, a non negligible fraction of a cluster's virial radius. \\

Importantly, most of the percentile intervals of the unrelaxed sample are about a factor of 2 larger than those of the relaxed clusters. 
Even for the BCG method, the 99-th percentiles increase drastically
from $\pm 19$ kpc for the relaxed clusters to asymmetrical extreme estimates of $(-684, +1570)$ kpc for the unrelaxed clusters.
Again, these values occur because there can be several DM peaks of
similar density due to subclusters located far apart from one another.
The finite number of projections, combined with the substructures, cause 
the 95-th and 99-th percentile tails of $\Delta y_{\rm BCG}$ of both the full
sample and the unrelaxed sample, but not the relaxed samples, to exhibit 
noticeable asymmetry.



\begin{figure}
\begin{center}
	\includegraphics[width=0.85\linewidth]{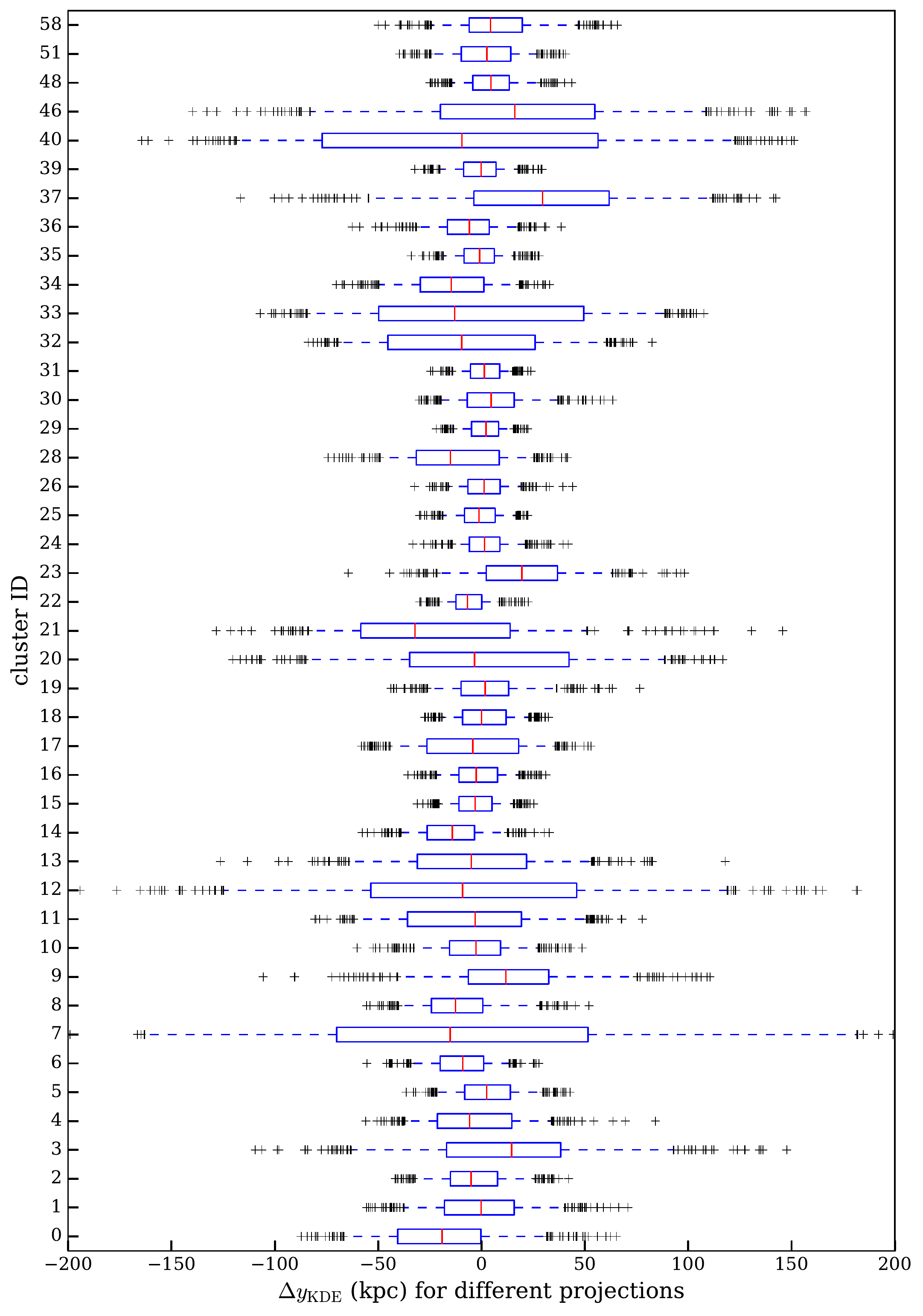}

	\caption{ 
	   Distribution of $\Delta y_{\rm KDE}$ for each cluster, across 768 projections per cluster. 
		The offsets are computed between the closest DM 
		peak to the brightest luminosity peak of each cluster. 		
		The red line of each box shows the median of the projections,
		the box encompasses the 25\% and 75\% percentile of the distribution while
		the whiskers mark the 5\% and the 95\% percentile. The other black crosses
		are data points with extreme values beyond the 5\% and 95\% percentile.
		The median of $\nu$ shown in Fig. \ref{fig:nu_distribution} and the max($\Delta y_{\rm KDE}$) values here show a correlation as high as 0.77
		\label{fig:projected_KDE_offset_distribution}.
	}

\end{center}	
\end{figure}

\subsection{Offset projection uncertainty of each cluster}
\label{subsec:projections}
As we gather the offsets in 768 projections for each cluster, we can estimate how much projection effects affect the uncertainties in the galaxy-DM offsets. We do so for the luminosity weighted KDE method. Distributions for each cluster across projections are given in the box plot of Fig.\ref{fig:projected_KDE_offset_distribution}. The values of the biweight 
mid-variance of $\Delta y_{\rm KDE}'$ for half of the clusters
are $< 23$ kpc. Of the ten clusters (ID = 3, 7, 12, 20, 21, 32, 33, 37, 40 and 46) that have mid-variance $ > 40$ kpc, all of them have the median of $\nu > 1.2$
over different projections. Indeed, the offset bias associated with projection effects is more extreme for the unrelaxed clusters.

\begin{table*}
	\centering
	\caption{Correlations between different cluster properties and the spread of the galaxy-DM offset obtained with luminosity weighted KDE method. See Section
		\ref{subsec:correlations} for details.
	\label{tab:correlations}} 
\begin{tabular}{lcccc}
\hline 
Significance & Quantity 1 &  Quantity 2 &  Correlation & Relevant section\\
\hline
High & unrelaxedness$_0$ & unrelaxedness$_1$ & 0.82 & \ref{subsubsec:relaxedness}
\\
& unrelaxedness$_0$ & $\max(\Delta y_{\rm KDE}')$ &  0.70 &
\ref{subsubsec:relaxedness}, \ref{subsubsec:KDE}
\\ 
& unrelaxedness$_1$ & $\max(\Delta y_{\rm KDE}')$ &  0.80 & \ref{subsubsec:relaxedness}, \ref{subsubsec:KDE} \\ 
& median($\nu$) &  $\max(\Delta y_{\rm KDE}')$ & 0.77 & \ref{subsubsec:nu},
\ref{subsubsec:KDE}\\
\\ \hline
Medium & median($\nu$) & $M_{\rm FoF}$ & 0.28 & \ref{subsubsec:nu},
\ref{app:table_of_results}\\
\\ \hline
Low & richness & $\max(\Delta y_{\rm KDE}')$ & 0.21 & \ref{subsubsec:KDE}, \ref{app:table_of_results}\\  
& $M_{200C}$ &  $\max(\Delta y_{\rm KDE}')$ & -0.14 &
\ref{app:table_of_results}, \ref{subsubsec:KDE}\\ 
& $M_{500C}$ & $\max(\Delta y_{\rm KDE}')$ & -0.18 &
\ref{app:table_of_results}, \ref{subsubsec:KDE}\\
& $M_{FoF}$ & $\max(\Delta y_{\rm KDE}')$ & 0.13 & \ref{app:table_of_results}, \ref{subsubsec:KDE} \\
\hline
\end{tabular} 
\end{table*}

\subsection{Correlations between offsets and cluster properties}
\label{subsec:correlations}

Here we systematically investigate how various physical properties of the simulated clusters correlate with the spreads of the offset distributions. We focus again on the offsets from the luminosity weighted KDE method. The considered cluster properties are listed in Table \ref{tab:correlations}, ranked in terms of the
significance of their correlation with a measure of $\Delta y_{\rm KDE}'$. 

We use the Pearson product-moment correlation coefficient to quantify linear 
relationship between the pairs of variables
(aka Pearson's r,  hereafter $\rho$).
We describe the significance of the correlation 
based on the p-value reported by {\sc{Scipy}} of seeing the level of 
correlation by chance assuming the pair of 
quantities has no correlation. As a reference, the correlation $\rho$ between the 
two unrelaxedness criteria defined in Section \ref{subsubsec:relaxedness}
for the 43 selected clusters is as high as 0.82.
 If the p-value is greater than 0.1, we consider the
correlation insignificant. Note that most physical quantities are unique per cluster, such as the mass or
the richness. For quantities that are not unique per cluster, but are projection-dependent,
such as the projected offset and the total normalized peak density ($\nu$), 
we compute a summary statistic, such as the
maximum of the offset, and the median of $\nu$, over all the projections of
each cluster.  Then we compute the correlation between these summary statistics and the other physical quantities that
are unique for each cluster.

Table \ref{tab:correlations} demonstrates that the projected offsets strongly correlate with the unrelaxedness of the clusters under study, either measured via the theoretical unrelaxedness criteria of Section \ref{subsubsec:relaxedness} or via the total normalized KDE peak density $\nu$ of Section \ref{subsubsec:nu}. No other cluster property, such as richness or mass, exhibits significant correlation with the spread of the projected offset, within the values encompassed by our selected clusters.

\section{Implications of our findings \& Discussion}
\label{sec:discussion}

We conclude our study by putting our Illustris findings into the broader context.
In Section \ref{subsec:non_SIDM_offsets}, we show how our offset distribution results compare with studies that investigate causes of offsets that do not have a SIDM origin. We compare Illustris offset levels to SIDM staged simulations results in Section \ref{subsec:SIDM_sim}. Finally and more importantly, in Section \ref{subsec:SIDMimplications} and \ref{subsec:prospect_of_detecting_SIDM} we discuss the implications that our CDM findings have in relation to the prospect of constraining SIDM via measurements of galaxy-DM offset in observed merging clusters.

\subsection{Illustris offsets in context: general studies of clusters and groups}
\label{subsec:non_SIDM_offsets}

The results presented above relative to the galaxy-DM offset distributions are highly relevant to two types of optical studies of galaxy clusters that do not necessarily pertain to SIDM constraints.
The first type includes lensing studies that aim to estimate the spatial maps of 
the mass of the DM distribution of galaxy clusters. 
The second type focuses on inferring cosmological parameters from the 
mass function of galaxy groups. To do so, galaxy groups are stacked to achieve a high enough signal for mass inference, and stacking requires a robust estimate of galaxy group centers. 

In Section \ref{sec:results}, we have demonstrated that the the BCG method has the tightest offsets from the DM peaks in cosmological simulations, followed by the KDE luminosity peak. 

These results are consistent with other theoretically-derived results. For example, \cite{Cui2015} have studied 184 galaxy clusters with M$ > 10^{14}$ M$_\odot$ in an
N-body and hydrodynamical cosmological simulation suite performed with {\sc GADGET-3}, and found the majority of offsets between BCGs and the most gravitationally bound particle to be below 10 $h^{-1}$ kpc. They also reported some extreme outliers 
spanning up to several hundred $h^{-1}$ kpc due to the disturbed morphology of
some clusters. Our tight 68-th percentile of 
$\Delta y_{\rm BCG}$ at $ \pm 3$ kpc gives some confidence that 
we have identified most of the BCGs correctly in the Illustris simulation.

However, the distributions of $\Delta s_{\rm BCG}$ derived from simulations are, in general, tighter than those measured in observations.
For example, \cite{Oguri2010} have analyzed 25 X-ray luminous 
massive galaxy clusters of the LoCuSS survey, using the Subaru Suprime Camera with a large FOV ($\sim 3 h^{-1}$ Mpc on a side).
By fitting elliptical NFW models to the weak lensing data, \cite{Oguri2010}
showed a long tail distribution for $\Delta s_{\rm BCG}$, which they fit with two 2D Gaussians.
The first had a standard deviation being 90
$h^{-1}$ kpc, a substantially larger width than we found for the
68-th percentile of $\offset_{\rm BCG}$. 
The long tail of the offsets from \cite{Oguri2010} spanning around 1 Mpc was fit by a second 2D
Gaussian with a standard deviation of 420 $h^{-1}$ kpc. This second component
in the tail region contains $\sim$ 10\% of the clusters in the study and is
consistent with the portion of extreme outliers that we have.   
Two major sources of uncertainty that do not depend on SIDM have been identified to justify the observed spreads of $\Delta s_{\rm BCG}$: misidentification of the BCG and details of the lensing modeling to identify the DM peak. 

To see the effects of BCG misidentifications, N-body cosmological
simulations have been used to estimate the 2D distances between the BCG and 
the second most massive galaxies. \cite{Johnston2007b} and 
\cite{Hilbert2010} (using the Millennium simulation)  found 
the one sigma level offsets for misidentified BCGs at 380 $h^{-1}$ kpc, and 410 
$h^{-1}$ kpc respectively. This is consistent with the 95-th percentile of the
unrelaxed and the full sample of the BCG in our study and also the tail of the
$|\Delta s_{\rm BCG}|$ for \cite{Cui2015}. If one wishes to use the BCG with high
confidence, it may be necessary to set a stringent standard on the morphological
characteristics such as requiring a large half light radius for classifying a BCG.

To isolate the effects of the lensing modeling, \cite{Dietrich2012} performed an analogous analysis of the work of \cite{Oguri2010} 
using the N-body Millennium Run (MR) simulation. By ray-tracing through 512 mocked cluster-sized haloes, \cite{Dietrich2012} showed that a combination of shape noise, smoothing bandwidth, number density of the source galaxies, and modeling choices can lead to hundred-kpc-level offsets between the most bound particle 
(a proxy of the BCG) and the lensing peak, while without any smoothing or shape noise, in 90\% of the cases the distance between the lensing peak and the 
most bound particle is around 2.0 $h^{-1}$ kpc
(0.65 arcsec at z = 0.3). 

Finally, additional observational analyses support our findings about the potential of the KDE luminosity peak to summarize the galaxy population of a cluster, especially in those cases with several bright galaxies that have about the same brightness 
in a cluster (or a subcluster) and no unique BCG can be identified. 
For example, \cite{George2012a} examined 129 X-ray selected non-merging galaxy 
groups in the COSMOS field and found that around 20\% to 30\% of groups have non-negligible discrepancies
between different galaxy centroids. 
By stacking on a bright galaxy near the X-ray centroid, they found  
the resulting lensing strength is higher than the stacked lensing signal based
on other galaxy centroids, including the BCG. 
For groups with clear BCG candidate, \cite{George2012a} gave the range of
offset between the BCG and the assumed halo center as $\lesssim 75$ kpc. 
The KDE peaks from the luminosity maps of the Illustris samples on the other hand show a much 
tighter offset to the 
lensing center than any other investigated galaxy summary statistics. 
The weighted or unweighted centroid measurement from \cite{George2012a} has a 
$|\Delta s|$ with standard deviation at 50 - 150 kpc from the
lensing center with long tails (of around several hundred kpc). 
In comparison, the median (26 kpc), mean (37 kpc), standard deviation (35 kpc) 
and 75-th percentile (49 kpc) of 
$|\Delta s_{\rm KDE}|$ from all the Illustris samples are below 50 kpc.

\subsection{Comparison to staged simulations with SIDM}
\label{subsec:SIDM_sim}
Staged simulations are controlled probes for the contribution
of SIDM to offsets in mergers of cluster components. 
The non-deterministic nature of particle interactions means that it is not easy
to predict the offsets analytically without simulations.
Furthermore, there is no consensus as to what type of clusters might best show the
effects of SIDM, as this may depend on the SIDM model. 
Currently, the studies of SIDM in mergers have been restricted to those with isotropic
scattering and velocity-independent cross sections.   
The two main classes of SIDM that are studied include those with 
frequent, long-range interactions that can be broadly be modelled by an
effective drag force, and those with rare, shorter range interactions those
effects are not well approximated by a drag-force \citep{Kahlhoefer14}. 

At SIDM cross section level favoured by current literature $\sigmaSIDM
\lesssim 1~\centi\meter^2 ~\gram^{-1}$, a list of SIDM simulation studies (\citealt{Kim:2016}, \citealt{Robertson2016}, \citealt{Kahlhoefer14}, \citealt{Randall2008d})
have reported that, when the offset is observable, 
the maximum offset generally increases with $\sigmaSIDM$. 
A {\it maximum} offset of 40 kpc was reported for $\sigmaSIDM = 1
~\centi\meter^2 / \gram$ under high concentration of the DM component of the 
merging clusters. Increasing the cross section to $\sigmaSIDM = 3
~\centi\meter^2 / \gram$ only increased the {\it maximum} offset to 
approximately 50 kpc.

The maximum signal of $\offset_\SIDM = 50$ kpc is within 
the one-sigma level of the Illustris offsets 
inferred by the shrinking aperture method, the number density method, and the
centroid method. This maximum value is also within the two sigma value of the
luminosity weighted KDE offset and the BCG offset level and several times below
$\max(\Delta y_{\rm KDE}')$ and $\max(\Delta y_{\rm BCG}')$. 

In the terminology of Equation \ref{eq:signal_and_noise} and Section 1, this means that the noise could dominate the signal, with 
\begin{equation}
	n_{\rm CDM} > \offset_\SIDM.
\end{equation}
If we want to ensure a high signal to noise ratio for possible
$\offset_\SIDM$ detection, one should choose to use either the KDE or the 
BCG method for inferring the offset.

\subsection{Implications for SIDM observational contraints}  
\label{subsec:SIDMimplications}
  
The spread of the galaxy-DM offsets which we have measured with various methods from Illustris clusters provide a baseline reference for the magnitude of observational uncertainties, systematic biases, and statistical noise in a $\Lambda$CDM cosmology: these can all affect our ability of constraining $\sigmaSIDM$ from the observed galaxy-DM offsets in merging clusters.

\subsubsection{Observations of merging clusters for SIDM} 

In Table \ref{tab:offset_results}, we collect results from the literature of 7 merging clusters which have been adopted to constrain SIDM. We include a total of 15 measurements of the galaxy-DM offset, by indicating in each case the underlying adopted method (mostly KDE luminosity peak, KDE number density peak and BCG methods). For most of the observed cases, we obtain the estimates for the offsets from the contour plots and descriptions in the corresponding papers. Keeping in mind the complexity in comparing 1D and 2D offset measures (see Section~\ref{subsec:1d}), we make our best attempt to measure the spatial component of the
observed offset along the axis connecting subclusters, if they exist, and we denote such 1-dimensional measure $\Delta y_{\rm obs}$.
For comparison, we also report $\Delta x_{\rm obs}$, the offset approximately perpendicular to
$\Delta y_{\rm obs}$.
Among the observed clusters, Abell 3827 is the only case with no
subclusters, although it has four bright galaxy peaks in the central region.   
For the offsets that are roughly in line with the axis connecting the two subclusters,
we let $\Delta y = |\Delta {\bf s}|$. 

As can be seen, observed 1D offsets range between a few kpc to up a few hundreds kpc, for merging objects with inferred mass between about $10^{14}$ and $10^{15} M_\odot$.  

\begin{table*}
	\centering
	 \caption{Observed offsets from clusters with reported evidence of mergers
		 along line connecting two subclusters ($\Delta y_{\rm obs}$) and the approximate 
		 perpendicular offset ($\Delta x_{\rm obs}$). 
		 The Table, but for the last column, is a collection of data from the literature and mainly contains clusters that have been used to constrain
		 $\sigmaSIDM$ using the reported offsets.
	Approximate  
	 error estimates are given as the 68\% lensing peak uncertainty
	 from the figure(s) listed in the reference column, as no other uncertainty estimates of the galaxy summary statistics are available in most of the literature. 
		Error estimates are omitted when they are not reported by the authors in
		any form. All masses are reported for the subclusters listed under the
		subcluster column. The last column reports the p-values (rounded as explained in the text) we find from comparing the observed offsets to the distribution of offsets measured from Illustris cluster data with the same definition of galaxy summary statistics. We find that 13 of the 15 reported offsets are consistent at the 95\% level with the offset distributions measure from Illustris $\Lambda$CDM clusters.
	 \label{tab:offset_results}} 
	  \begin{tabular}{@{}lccccccccc@{}}
	 \hline 
	 Cluster &$\Delta y_{\rm obs}$ & $\Delta x_{\rm obs}$ & $|\Delta s_{\rm obs}|$ & galaxy peak & DM peak &  subcluster &  mass &  reference & p-value\\
	 &(kpc) &(kpc) &  (kpc) &  &  &   &$10^{14}$ M$_\odot$ &   & \\
	 
	 \hline
	 Bullet  & 9  & -23 & 25$~\pm~29$ & num. or lum. & SL \& WL  & northwest & 1.5 & \citealt{Randall2008d}& $0.32$\\
	 Baby Bullet & -40&  0 & $\sim 40 \pm \sim 50 $  & lum. & SL \& WL & northwest & 2.6 & \citealt{Bradac2008}:Fig.4 & $ 0.05 $\\
	 Baby Bullet & 30  & 0 & $\sim 30 \pm \sim 75 $ & lum. & & southeast  & 2.5 & \citealt{Bradac2008}:Fig.4 & $ 0.32 $\\
	 Musketball &  129 & 0 & 129 $\pm \sim63$ & num. & WL  & southern & 3.1 & \citealt{Dawson2013}:Fig.4.7 & $0.05$\\
	 Musketball & -47 & 0 & 47 $\pm \sim50$ & num. & & 	 northern & 1.7 &  \citealt{Dawson2013}:Fig.4.7& $0.32$ \\
		Abell 3827 & 6 & 0 & 6 & BCG & SL  & central & & \citealt{Williams2011a}& $0.05$\\ 
		Abell 520 & 0 & 50& $\sim50 \pm \sim50$ & lum. & WL  & blue & 5.7 & \citealt{Clowe2012}:Fig. 4 & $0.32$\\
		El Gordo &  58 &0 & $\sim58 \pm \sim100$ & lum. & WL & 	northwest& 11  &\citealt{Jee2014}:Fig.7,8  & $0.05$ \\
		El Gordo & 30 & 110 & $ 115 \pm \sim60$ & num. & & northwest&   &\citealt{Jee2014}:Fig.7,8  & $0.32$ \\
		El Gordo & 6 & 25& $\sim26 \pm \sim50$ & lum. & & northwest & 7.9   &\citealt{Jee2014}:Fig.7, 8  & $0.32$ \\
		El Gordo & 280 & 280 & 400 $\pm \sim40$ & num. &  &southeast &   &\citealt{Jee2014}:Fig.7, 8  & $0.05$\\
		Sausage &160 & 100& $\sim190\pm \sim150$ & num. & WL  & north & 11.  &\citealt{Jee2015}:Fig.10 & $ 0.05$\\ 
		Sausage &160 & 160& $ \sim190\pm \sim150 $  & num. &  & south & 9.8 & \citealt{Jee2015}:Fig.10 & $ 	0.05$ \\ 
		Sausage & 320 & 130 & $\sim340 \pm \sim150 $  & lum. &  & north & 11. & \citealt{Jee2015}:Fig.10 & $\lesssim 0.01$\\ 
		Sausage & 160 & 160 &$\sim230 \pm \sim150 $  & lum. & & south & 9.8 &\citealt{Jee2015}:Fig.10& $\lesssim 0.01$\\ 
	 \hline
	 \end{tabular}

	 \raggedright{
		 {\it num.} is a short hand for the KDE peak estimate from the galaxy number
		 density map. \\
		 {\it lum.} is a short hand for the KDE peak estimate from the luminosity
		 density map, or KDE' in the method description. \\
		 {\it SL} is a short hand for strong lensing. \\
		 {\it WL} is a short hand for weak lensing. \\
	 }
\end{table*}

\subsubsection{Constructing the hypothesis test: p-values from the null hypothesis} 
\label{subsec:hypothesis_test}

We get an estimate of the actual significance of each observation by comparing the
observed offsets to the distribution of $\Delta y$ measured from Illustris cluster data. 
The distributions of $\Delta y$ represent the possible ways that offsets can be observed in a CDM
universe, giving us a rough estimate of the probability 
of seeing the offset from observations under the null hypothesis of CDM 
being true. 
We therefore compute the two-tailed 
p-value from the narrowest density interval (C) of simulated offsets 
that is above the observed values of offsets in the literature, 
i.e. the significance of each observed offset is rounded up to the nearest 68\%, 95\% or 99\%
interval of the corresponding offset distribution from the Illustris data:
\begin{equation}
	p = 1 - C(|\Delta y| > \Delta y_{obs})
\end{equation}
This underestimates the probability that the
observed offset is compatible with the CDM model so
any disagreement with the CDM model will be more obvious. 
For each observed case, the comparison is performed by considering the distribution of Illustris $\Delta y$ for the unrelaxed cluster sample ($1.2 < \nu < 2.2$) and obtained with the same observationally-adopted methods.

We find that 6 of the 15 observed offsets have p-values lower bounds of 0.32, 7 of them have p-values $ \ge 0.05$, and only 2 p-values are below 0.01. The latter are the offsets measured in the Sausage cluster with the luminosity density peak galaxy summary statistics; for the same object, \cite{Jee2015} have found p-values $<$ 0.288 and 0.082 for the two substructures, respectively, in comparison to their evaluated level of noise. 
Here, based on our {\it optimistic} estimate of the level of noise affecting the measurement of the galaxy-DM offsets in a CDM scenario, we find that 13 out of the 15 observed galaxy-DM offsets are consistent at the 95\% level with the hypothesis that $\Lambda$CDM is the true underlying physical model.
This means that the observations are mainly consistent with the null hypothesis, namely it is possible to see offset values as extreme as reported by observations in a CDM universe $n_{\rm CDM}$, i.e. 
\begin{equation}
\offset_{\rm obs} \approx n_{\rm CDM}. 
\end{equation}
\subsubsection{Limitations of the current Illustris-based analysis}
\label{subsec:limitation_of_pvalue}

We have just shown that most individual offsets observed from merging clusters are consistent or marginally consistent with offset distributions drawn from somewhat idealised mock observations of $\Lambda$CDM simulated clusters. 
In fact, more realistic modelling of the observational process may increase the simulated LCDM offsets even further.

Throughout this paper, we have indeed aimed at quantifying the optimistic noise floor in the measurements of the galaxy-DM offsets and hence, by design, presented the results from Illustris cluster offsets in ideal observational conditions. There are therefore reasons to believe that the p-values quoted above may be underestimated, as we expect more realistic observation models to substantially widen the noise ($n_{\rm CDM}$) distributions.


First, we have assumed an unobstructed line-of-sight for our clusters. 
In real observations, there can be foreground galaxies, whose noise contribution
theoretically has a uniform spatial distribution. 
This contribution can add large spatial variance to the estimate of ${\bf
s}_{\rm gal}$. 

Second, our DM mass map has assumed a resolution that is only achieved for clusters 
with many strong lensing constraints. 
Most analyses of clusters have DM maps with lower resolution. 
The observed galaxy clusters can also have high stellar contamination
if the cluster happens to be located near the plane of the Milky Way 
(such as the Sausage cluster; \citealt{Jee2015}). 
Such stellar contamination can decrease the number density of the visible source galaxies for the
lensing analysis, resulting in a degraded resolution of the DM mass map and a
biased estimate of ${\bf s}_{\rm DM}$. 
  
Third, most studies have {\it not} used estimators of comparable precision, especially for
the estimate of the luminosity peak. We have used the cross-validated KDE that
minimises the fitting error. Not only does the algorithm help
determine the eigenvalues of the bandwidth matrix for smoothing, 
but also the optimal eigenvector direction of the matrix. 
Most literature does not treat the inference of the smoothing bandwidth 
of the data to be a regression problem that aims to minimize the fitting error.  
Doing so
avoids setting the bandwidth to fit the preconception of how the density
contours of the cluster 
should look like, and inadvertently biasing $\Delta s$. We therefore strongly
advocate treating the smoothing of galaxy luminosity as a regression problem.
The best fit bandwidth should be found either via 
the cross-validated KDE method that we provide, or other optimization  
procedures that minimize the fitting error.


Other complications may arise from the mismatch of the physical properties of 
our cluster samples with the observed clusters, although the actual effect of the
property mismatch is inconclusive. 
For the full sample of Illustris clusters in Table \ref{tab:p_val_table}, 
each cluster has the same number of projections. 
However, it also underestimates the offset spread because the
full sample includes $\sim 45\%$ of relatively relaxed projections 
that only have one primary luminosity component.  These clusters would
have been excluded for comparison with bimodal mergers. 
Subsetting with $1.2 < \nu < 2.2$ picks out
cluster projections that are in more similar dynamical states as the observed merging
clusters. 
However, some simulated clusters may have more projections included in this sample
than the other clusters. From the inspection of the mass abundance relations in 
Fig. \ref{fig:mass_abundance_distribution}, we found that subsampling with $1.2 <
\nu <2.2$ includes a higher proportion of projections from massive clusters
($\sim 20\%$ more) than 
the full sample. 
This sampling should not introduce significant bias, 
since the sample of observed merging clusters are still more massive than our
unrelaxed samples. The weak correlation value between the $M_{\rm FoF}$ and
$\Delta y_{\rm KDE}'$ implies that using less massive clusters may not lead to a significant
change of our estimation of $n_{\rm CDM}$.  
%
%
%
%
%
Due to the above possible discrepancies, we do not provide conclusions by combining the p-values from
the observations, as the computation of each p-values does not fully
take the uncertainties of the observations into account. 
Any future studies that wish to claim significance based on a 
p-value comparison to a simulation will 
need to carefully consider the contribution of uncertainties 
from each aspect of the cluster analysis. 


\subsection{Prospects of detecting SIDM} 
\label{subsec:prospect_of_detecting_SIDM}

In this paper, we have presented the results of a null hypothesis test aimed at quantifying the level of observational uncertainties, systematic biases, and statistical noise that arise in the measurement of the galaxy-DM offsets in clusters in a CDM scenario. In practice, following eq. \ref{eq:signal_and_noise} which here we repeat for convenience:
\begin{equation}
\nonumber
	\offset_{\rm obs} = \offset_{\SIDM} + n + \cdots,
\end{equation}
with our results, we have provided an optimistic estimation of a noise term $n_{\rm CDM}$ arising in a CDM scenario.
 
How can we improve upon the results presented thus far in order to constrain SIDM? We argue that a more robust statistical model is needed in order to properly take into account all possible sources of offset that can contribute to the observed ones, as per equation above.

It would be more appropriate to treat the $\offset_\SIDM$ term as a distribution. This follows from combined results from staged simulations \citep{Kahlhoefer14, Kim:2016}: $\offset_\SIDM$ varies according to the cluster merger parameters (merger
velocity and impact parameter), the cluster properties (spatial distribution of mass), and the merger phase (time dependence). Hence such variations of $\offset_\SIDM$ can be represented with a distribution.


In such a framework and assuming that the SIDM signal has little-to-no correlation with the noise, we can evaluate the discrepancy to be expected between a CDM model 
and a SIDM model by decomposing the variance as:
\begin{align}
	\Var(\Delta y_{\rm obs}) &= \Var(\Delta y_\SIDM) + \Var(n_{\rm CDM}) + \cdots
\end{align}

For example, a rough estimate of the distribution of $\offset_\SIDM$ over time 
can be read off from the staged simulation of an equal mass cluster merger of a
massive cluster ($10^{15}$ M$_\sun$) in \cite{Kim:2016}: an approximately zero mean and a 3-sigma level of
$\sim 40$ kpc for $\sigma_\SIDM = 1~\centi\meter^2 / \gram$.  
We can assume a noise level as estimated by the distribution of $\Delta y_{\rm KDE}'$
for the Illustris non-relaxed clusters (one-sigma level at 32 kpc - a choice that matches 
how \cite{Kim:2016} used the DM and the galaxy peaks for their offset measurements). 
Then a rough estimate of the difference in the standard deviation of the observed offset 
between a CDM model and a model with $\sigma_\SIDM = 1~\centi\meter^2 / \gram$ is:
\begin{equation}
\sqrt{\Var(\Delta y_\SIDM) + \Var(n_{\rm CDM})} - \sqrt{\Var(n_{\rm CDM})}  \approx 3~\mathrm{kpc} 
\end{equation}
which is a subtle $\sim 9 \%$ difference between the CDM and the SIDM model with
$\sigma_\SIDM = 1~\centi \meter^2 / \gram$. 
Note that the assumed values of $\Var(\Delta y_\SIDM)$  
ignore the possibility
that future studies may find ways to select mergers at a stage when $\Delta y_\SIDM$
of each cluster merger is around its maximum possible value. 

We can also estimate the difference of the standard deviation of the observed
offset level between two models with 
$\sigma_\SIDM =1~\centi\meter^2 / \gram$ and  $\sigma_\SIDM = 3~\centi\meter^2 / \gram$, respectively.
By assuming an approximately zero mean and a 3-sigma level of $\Delta y_\SIDM$
at $\sim 50$ kpc for $\sigma_\SIDM =
3~\centi\meter^2 / \gram$, we get

\begin{dmath}
\nonumber
	\sqrt{\Var_{3\centi\meter^2/\gram}(\Delta y_\SIDM) + \Var(n_{\rm CDM})} \hfill -
	\nonumber
	 \sqrt{\Var_{1\centi\meter^2/\gram}(\Delta y_{\SIDM}) +\Var(n_{\rm
	CDM})}\label{eq:variance_difference} \approx 1~\mathrm{kpc} 
\label{eq:variance_difference} 
\end{dmath}

which is an even smaller $\sim 3\%$ difference between the two models. 
If the distribution of noise from observations has larger variance than what we
have shown, then equation above will show even less discrepancy
between the different SIDM models and the offset distribution with models of small $\sigma_\SIDM$ values 
can be indistinguishable.  

For future studies, there may be other ways to better formulate the
statistical framework to constrain SIDM. For example, an alternative to the variance could be to compute a Bayes factor based
on all the important contributions to the observed offset. This would read as the ratio of the posterior probability of a fit of SIDM model to data,
over those of a CDM model:
\begin{equation}
	\tau = \frac{Pr(\sigma_{\SIDM} = \sigma | \offset_{\rm
	obs})}{Pr(\sigma_{\SIDM} = 0 | \offset_{\rm obs})}.
	\label{eq:Bayes_factor}
\end{equation}
This ratio will show if a SIDM model with an assumed cross section value, $\sigma$,
is favored over a CDM model, by providing better
insight to the best-fit $\sigma_\SIDM$ value than our hypothesis test.
Our current hypothesis test can only quantify the compatibility between a CDM
model and the data but it does not describe the SIDM model. 

However, limitations come from many sides. On the one hand, there are not enough observational data points for us to estimate 
$\Var(\Delta y_{\rm obs})$ to a high precision. In fact, the current methods for inferring 
the galaxy peaks may not be precise enough to achieve the optimistic noise level we have found in this study. On the other side, we do not know $\Var_{\sigma_\SIDM}(\Delta y_\SIDM)$ well enough
to have meaningful constraints: a better estimate would require ever more sophisticated theoretical models in SIDM scenarios which encompass cluster configurations analog to the observed ones. 
Finally, we would need to devise selective observational and modeling strategies to only study clusters that have a good chance of giving high $\offset_\SIDM$ signal
relative to the noise. For example, \cite{Kim:2016} found that such high SIDM-signal clusters are massive cluster mergers with relatively low merger
velocities, small impact parameters, and large halo concentrations. 
Under such restrictive selection criteria for cluster samples, not even the Illustris contain many analogous clusters for computing the denominator term in
\ref{eq:Bayes_factor}, and larger cosmological volumes or statistical samples of CDM cluster models are needed.





\section{Summary and Conclusions}
\label{sec:conclusions}

We have used Illustris, a large-scale hydrodynamical simulation in $\Lambda$CDM cosmology, with two purposes:
a) to examine the accuracy of commonly adopted techniques to quantify the galaxy-DM offset in galaxy clusters; and hence b) to quantify the implications our results have for studies that make use of the observed offsets from merging clusters to constrain self-interacting dark matter, $\sigma_\SIDM$.
In particular, we have measured the spatial offset between \\

1) the {\it peak} or summary statistics of the dark matter as traced by its smooth and clumpy components and typically observationally probed via gravitational lensing measurements; and \\

2) the {\it peak} or summary statistics of the discrete spatial distribution of cluster member galaxies and observationally probed with optical imaging.\\

By assuming $\sigma_{\rm SIDM} = 0$ and hence vanishing galaxy-DM offsets, we have used the Illustris simulation to quantify how likely it is for us to see the galaxy-DM offset values observed in merging clusters of galaxies.

To this aim, we have selected Illustris clusters and groups at $z=0$ that contain at least 50 member galaxies in turn exceeding an observationally-motivated magnitude cut of $i \leq 24.4$ (apparent at $z\sim0.3$). This selection resulted in 43 galaxy clusters with masses in the range $\sim 10^{13} - 1.6 \times 10^{14}$ M$_\odot $ and including a variety of evolutionary and relaxedness stages: we have studied them in 768 projections per cluster and encompassing fields of view dictated by the {\sc FoF} algorithm, i.e. usually $\sim $1 Mpc per side. 

We have used the galaxy cluster members characterised by their $i$-band luminosity to measure commonly-used galaxy summary statistics: kernel-density-estimation (KDE) luminosity peak, KDE number density peak, shrinking aperture method, centroid and position of the brightest cluster galaxy (BCG). We have used the dark-matter particles as tracers of the underlying DM density to replicate commonly-adopted methods to find the DM summary statistics, i.e. typically peaks of the gravitational lensing maps. 

First of all, the KDE method has provided us with an observationally-reproducible way to characterise the amount of substructures in a cluster map, i.e. its {\it relaxedness}. With high completeness and purity of member galaxy data, the luminosity maps produced by a cross-validated kernel density estimate resemble the DM spatial distribution more closely than the number density map of member galaxies: the normalised sum of the relative densities of the luminosity peaks in a map ($\nu$) correlates well with theoretical definitions of relaxedness. This has allowed us to divide our simulated cluster projections into a relaxed ($\nu < 1.2$) sample and a not-relaxed ($1.2 < \nu < 2.2$) sample, the first indeed characterised by a dominant luminosity peak and negligible substructures, the latter including about 50\% of the projections. \\

From our comprehensive study of the various ways of summarizing 
galaxy-DM offset using the Illustris simulation, we conclude the following:

\begin{itemize}

		\item The BCG has the smallest one-sigma offset level to the dominant DM
			peak (68-th percentile of $\Delta y_{\rm BCG} \approx 3$ kpc for unrelaxed clusters: $1.2 <
			\nu < 2.2$).\\

		\item Yet, the identified BCG offsets exhibit a 5\% tail at $\Delta y_{\rm BCG} >
			160$ kpc for $1.2 < \nu < 2.2$. 
		 This heavy tail is due to a combination of effects from substructures and 
		 projection and is not seen in the relaxed sample with $\nu < 1.2$. \\

		\item The KDE peak of the luminosity map after careful cross-validation 
			gives the second tightest one-sigma offset level ($\pm 32$ kpc for $1.2 <
			\nu < 2.2$) from the DM peak.  
			The length of the tail of the $\Delta y_{\rm KDE}'$ distribution
			($\lesssim \pm 124$ kpc at the 99-th percentile)
			is smaller than that of the BCG.\\

		\item A naive implementation of the shrinking aperture is easily affected 
			by substructures even for clusters with one
			dominant component. We do not endorse this method for drawing scientific
			conclusions nor visualization.\\

		\item The variances of the galaxy-DM offset measured from different methods
			of inferring the galaxy peak can differ by factors of a few (Table \ref{tab:p_val_table}). 
			Hence, one should not compare offset values computed with different galaxy-statistics methods.
			
\end{itemize}

Our results on the galaxy-DM offset in a $\Lambda$CDM cosmology provide an {\it optimal} measure of the possible uncertainties and systematic bias which can affect the interpretations of observed galaxy-DM offsets in merging clusters for constraining $\sigma_\SIDM$.

Importantly, we have collected results from the literature of 15 observed offsets from merging clusters (see Table \ref{tab:offset_results}) and statistically compared the reported offsets with the ones from the Illustris unrelaxed sample measured with the same methods. We find that it is possible to see offsets as extreme as those observed in merging galaxy clusters by assuming that $\Lambda$CDM is the true underlying physical model, i.e. $\Var(\offset_{\rm obs}) \approx \Var(n_{\rm CDM})$.\\	

Moreover: 
\begin{itemize}

		\item The contribution of statistical uncertainty to the galaxy-DM offsets 
			for $\Lambda$CDM clusters is {\it not} negligible when compared to the reported  
			levels of maximum offset from staged SIDM simulations ($\sim 50$ kpc),
			i.e. $\Var(n_{\rm CDM}) \gtrsim \max(\Delta y_{\SIDM})$.\\
		
		\item To maximize the SIDM signal, one should use either the BCG or the
			luminosity weighted KDE methods for measuring the offset.
			For the non-relaxed samples in our study ($1.2 < \nu < 2.2$),
		  only the 68-th percentile of the galaxy-DM offsets derived from BCG ($y_{\rm BCG}$) and 
			the KDE luminosity peak	($y_{\rm KDE}'$)
			are smaller than the maximum SIDM offset ($50$ kpc) reported from SIDM staged simulations. 
			Other methods have one-sigma uncertainty levels that overwhelm the
			SIDM offset signal. 


\end{itemize}

Finally, we caution that this analysis has been carried out with the purpose of estimating the {\it minimum} level of statistical noise arising in a $\Lambda$CDM cosmology, and therefore by treating the Illustris simulated clusters in somewhat idealized mock-observation conditions. We therefore expect the width of the herein reported Illustris galaxy-DM offset distributions to be underestimated and to increase by including progressively more sophisticated and realistic models to reproduce observations. 

In conclusion, given the spread of the CDM galaxy-DM offsets and our quantitative findings, 
we advocate that, in order to improve the prospects of constraining SIDM, careful theoretical studies will need to be undertaken in order to identify the populations of merging clusters that maximize the information content relatively to SIDM observables and that more robust statistical models will need to be formulated in order to properly take into account contributions to the observed offsets which are not due to SIDM.





\section{Acknowledgements}
Karen Ng would like to thank Prof. Thomas Lee for the helpful discussion of 
the construction of the p-value hypothesis test. We are immensely grateful for 
the technical review and comments from Prof. Maru\v{s}a Brada\v{c} and Prof. Annika Peter.  
Part of the work before the conception of this paper was discussed during 
the AstroHack week 2014. Part of this work was performed under the HST grant
HST-GO-13343.01-A, the National Science Foundation grant 
1518246 and under the auspices of the U.S. Department Of Energy by 
Lawrence Livermore National Laboratory (LLNL) 
under Contract DE-AC52-07NA27344; \ifthenelse{\boolean{thesis}}{}{LLNL-JRNL-704920-DRAFT} AP acknowledges support from the HST grant HST-AR-13897.
This work made use of {\sc IPython}
\citep{Perez2007}.

\bibliographystyle{mnras}
\bibliography{galaxyDMoffset,Rpackage}

\appendix
\section{Algorithm of the Shrinking aperture estimates}
\label{app:shrink_apert}
\begin{algorithm}
	\caption{Shrinking aperture algorithm with luminosity weights}
	\KwData{subhalo that satisfy cuts as a galaxy}
	 \hrulefill \\

	 initial aperture centroid = weighted mean galaxy location in each spatial dimension\\
 	distance array = euclidean distances between initial aperture center and each galaxy
	location \\
 	aperture radius = 90th percentile of the weighted distance array\\ 
	\While{ (newCenterDist - oldCenterDist) / oldCenterDist $\geq$ 2e-2}{
 		new data array = old data array within aperture\\
 		newCenter = weighted mean value of new data along each spatial dimension 
	}   \hrulefill
\end{algorithm}

\section{Table of results}
\label{app:table_of_results}
In this Appendix we collect corollary results of our analyses, together with a table reporting the physical and offset summary statistics of all the 42 clusters selected from Illustris.

\begin{figure}
	\begin{center}
	\includegraphics[width=0.95\linewidth]{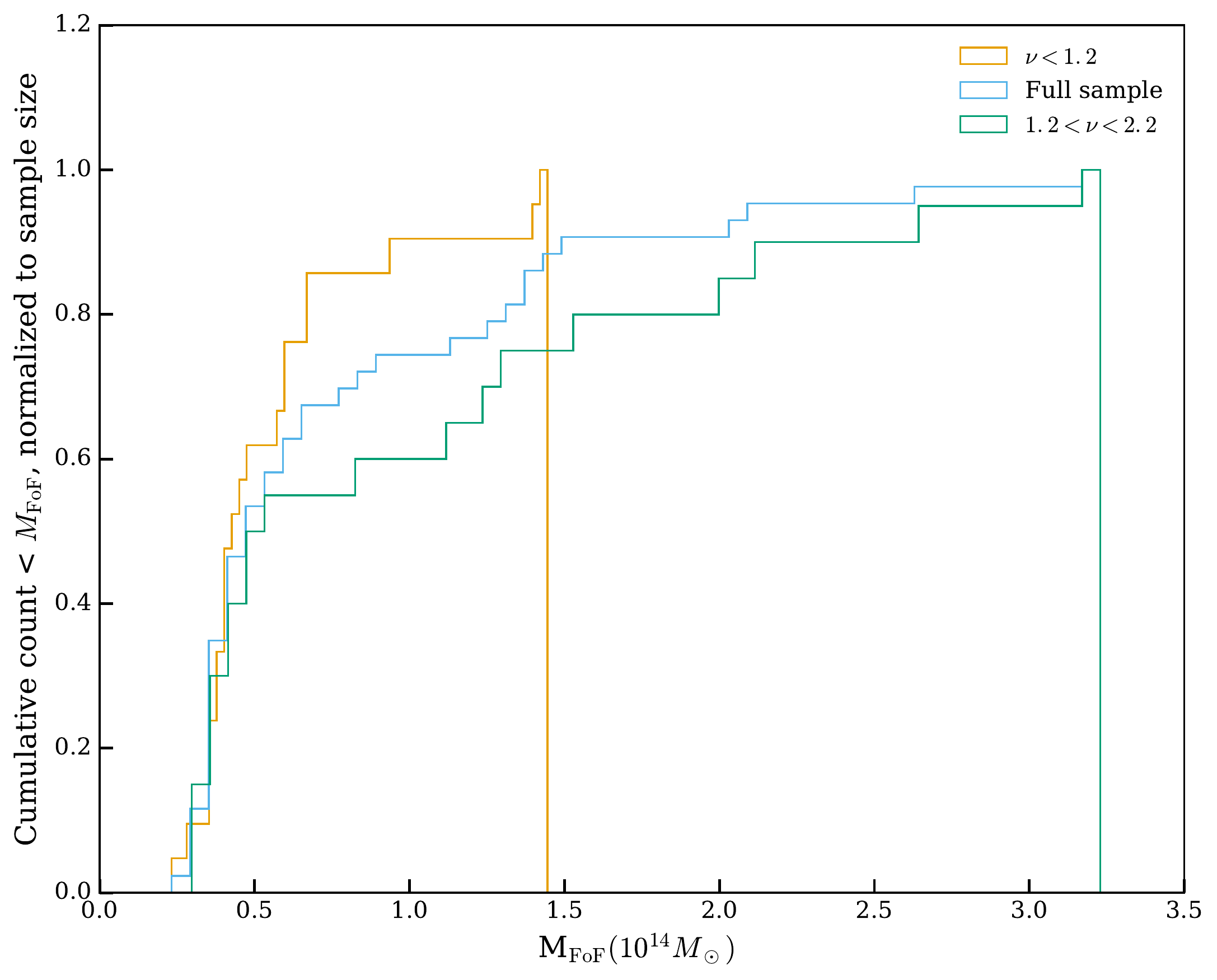}
	\caption{Cumulative distribution of Illustris clusters below a certain mass threshold
		for different samples: all selected clusters, relaxed ($\nu < 1.2$), and unrelaxed ($1.2 < \nu < 2.2$).
		Each distribution is normalized to the sample size.
		If the subsets have the same cluster mass abundance as the full sample,
		the three curves would like on top of one another. Instead, relaxed clusters appear only at the low mass end, while the unrelaxed ones dominate the sample for masses larger than $ \gtrsim 1.5 \times 10^{14}$ M$_\odot$.
		\label{fig:mass_abundance_distribution}
	}
\end{center}
\end{figure}

\begin{table}
	\begin{center}
	\caption{Properties of the offset distributions
		between the most bound particle and various summary statistics of 
		the member galaxy population.
	\label{tab:most_bound_particle_offset_distributions}}
\begin{tabular}{lccccccc}
\toprule
offset	      &  68\% 	&  95\% &  99\% &  68\% 	&  95\%  & 99\%  \\
kpc	      		& lower 	&  lower & lower &  upper 	&  upper & upper  \\
\midrule
$\Delta y_{\rm BCG}$       &       -2 &          -2 &        -252 &          2 &         528 &        1107 \\
$\Delta y_{\rm centroid}'$ &        -134 &        -491 &       -1176 &         134 &         491 &        1176 \\
$\Delta y_{\rm KDE}'$      &        -19 &         -82 &       -1182 &          19 &          82 &        1182 \\
$\Delta y_{\rm num.dens}$  &       -83 &        -302 &       -1114 &          83 &         302 &        1114 \\
$\Delta y_{\rm shrink}'$   &          -50 &        -288 &       -1025 &          50 &         288 &        1025 \\
\bottomrule
\end{tabular}
\end{center}
\footnotesize{The prime symbol $'$  denotes offsets measured by adopting some luminosity weighting for the galaxy 
data}
\end{table}

\begin{table}
	\begin{center}
	\caption{Properties of the offset distributions between the DM peak and the estimated galaxy summary statistics for Illustris clusters.
		All 43 clusters and all 768 projections are used in this table. Here we compare 1D with 2D offsets, following the same methodology adopted for the 1D results of Table \ref{tab:p_val_table}.
	\label{tab:offset_distributions}}
	\begin{tabular}{lrrrrrrr}
\toprule
kpc &  mean &  std &   min &  25\% &  50\% &  75\% &  max \\
\midrule
$|\Delta s_{\rm BCG}|$       &    69 &  294 &     0 &    2 &    3 &    7 & 2335 \\
$\Delta x_{\rm BCG}$         &   -14 &  226 & -2331 &   -2 &   -0 &    1 & 2327 \\
$\Delta y_{\rm BCG}$         &    23 &  197 & -1980 &   -2 &    0 &    2 & 2332 \\
$|\Delta s_{\rm centroid}'|$ &   261 &  209 &     2 &  114 &  202 &  317 & 1103 \\
$\Delta x_{\rm centroid}'$   &   -42 &  224 & -1022 & -164 &  -37 &   66 & 1101 \\
$\Delta y_{\rm centroid}'$   &     0 &  244 & -1102 & -111 &   -0 &  111 & 1100 \\
$|\Delta s_{\rm shrink}'|$   &   118 &  156 &     0 &   21 &   60 &  165 & 1454 \\
$\Delta x_{\rm shrink}'$     &    -7 &  131 & -1089 &  -39 &   -3 &   23 &  969 \\
$\Delta y_{\rm shrink}'$     &     0 &  145 & -1091 &  -32 &    0 &   32 & 1109 \\
$|\Delta s_{\rm KDE}'|$      &    37 &   35 &     0 &   14 &   26 &   49 &  498 \\
$\Delta x_{\rm KDE}'$        &    -2 &   35 &  -330 &  -17 &   -2 &   12 &  386 \\
$\Delta y_{\rm KDE}'$        &    -0 &   37 &  -439 &  -15 &    0 &   15 &  440 \\
$|\Delta s_{\rm num. KDE}|$    &   136 &  161 &     1 &   56 &   92 &  147 & 2126 \\
$\Delta x_{\rm num. KDE}$    &   -12 &  142 & -1967 &  -55 &   -4 &   53 &  993 \\
$\Delta y_{\rm num. KDE}$    &    -0 &  155 & -1415 &  -54 &   -0 &   54 & 1417 \\
\bottomrule
\end{tabular}

\end{center}
\footnotesize{The prime symbol  $'$ denotes offsets measured by adopting some luminosity weighting for the galaxy 
data}
\end{table}

\begin{table*}
	\begin{center}
	\caption{Properties of the galaxy clusters selected from the Illustris simulation and used in the analysis. Richness is
	computed based on $i-$band $< 24.4$ assuming $z=0.3$ (see Section \ref{sec:illustris_sim}) and the relaxedness criteria (columns 6 and 7) are defined in Section \ref{subsubsec:relaxedness}.\label{tab:cluster_prop} }
	\begin{tabular}{lccccccccc}
\toprule
ID & richness & M$_{\rm 200C}$ & M$_{\rm 500C} $ & M$_{\rm FoF}$  & unrelaxedness$_0$ & unrelaxedness$_1$ & midvar($\Delta y_{\rm KDE})$ & max($\Delta y_{\rm KDE})$& median($\nu$) \\
& & $(10^{14} M_\odot)$ & $(10^{14}M_\odot)$ & $(10^{14} M_\odot)$  &  & & (kpc) & (kpc) & \\
\midrule
 0 &      483 &                             1.64 &                             1.09 &                             3.23 &              29 &              33 &                                 31 &                              65 &       1.43 \\
 1 &      338 &                             1.57 &                             0.62 &                             2.68 &              20 &              16 &                                 25 &                              71 &       1.59 \\
 2 &      267 &                             1.53 &                             0.87 &                             2.12 &              17 &               3 &                                 18 &                              42 &       1.30 \\
 3 &      343 &                             0.82 &                             0.56 &                             2.03 &              37 &              59 &                                 44 &                             148 &       2.01 \\
 4 &      213 &                             1.19 &                             0.66 &                             1.54 &              21 &               4 &                                 24 &                              84 &       1.58 \\
 5 &      212 &                             0.90 &                             0.56 &                             1.44 &              20 &              27 &                                 16 &                              43 &       1.19 \\
 6 &      225 &                             0.96 &                             0.60 &                             1.40 &              18 &               7 &                                 15 &                              28 &       1.16 \\
 7 &      230 &                             0.31 &                             0.17 &                             1.41 &              54 &             280 &                                101 &                             379 &       2.83 \\
 8 &      148 &                             0.83 &                             0.54 &                             1.34 &              24 &              26 &                                 20 &                              52 &       1.32 \\
 9 &      187 &                             0.79 &                             0.50 &                             1.29 &              23 &              12 &                                 33 &                             111 &       1.45 \\
10 &      158 &                             0.73 &                             0.53 &                             1.15 &              19 &               8 &                                 19 &                              49 &       1.35 \\
11 &      134 &                             0.57 &                             0.33 &                             0.95 &              20 &               9 &                                 36 &                              78 &       1.12 \\
12 &      164 &                             0.20 &                             0.09 &                             0.87 &              64 &             142 &                                 77 &                             218 &       1.99 \\
13 &      115 &                             0.22 &                             0.14 &                             0.79 &              63 &             143 &                                 38 &                             118 &       2.26 \\
14 &       90 &                             0.45 &                             0.29 &                             0.67 &              15 &               8 &                                 17 &                              33 &       1.08 \\
15 &       92 &                             0.51 &                             0.35 &                             0.68 &              11 &               3 &                                 11 &                              25 &       1.00 \\
16 &      113 &                             0.40 &                             0.23 &                             0.61 &              19 &               4 &                                 13 &                              31 &       1.06 \\
17 &       97 &                             0.42 &                             0.18 &                             0.60 &              21 &               8 &                                 27 &                              53 &       1.09 \\
18 &       83 &                             0.45 &                             0.31 &                             0.59 &              15 &               8 &                                 14 &                              32 &       1.20 \\
19 &       86 &                             0.26 &                             0.19 &                             0.57 &              30 &              68 &                                 18 &                              77 &       1.52 \\
20 &       84 &                             0.15 &                             0.11 &                             0.50 &              60 &             122 &                                 54 &                             117 &       1.65 \\
21 &       89 &                             0.26 &                             0.12 &                             0.53 &              23 &               8 &                                 47 &                             146 &       1.24 \\
22 &       70 &                             0.42 &                             0.30 &                             0.49 &              14 &               7 &                                 10 &                              23 &       1.06 \\
23 &       68 &                             0.25 &                             0.17 &                             0.47 &              30 &              25 &                                 26 &                              98 &       1.03 \\
24 &       66 &                             0.33 &                             0.26 &                             0.44 &              14 &              14 &                                 11 &                              42 &       1.17 \\
25 &       79 &                             0.23 &                             0.15 &                             0.43 &              23 &              25 &                                 11 &                              22 &       1.25 \\
26 &       61 &                             0.26 &                             0.18 &                             0.45 &              28 &              40 &                                 11 &                              44 &       1.30 \\
28 &       69 &                             0.30 &                             0.16 &                             0.41 &              22 &              12 &                                 26 &                              42 &       1.01 \\
29 &       62 &                             0.30 &                             0.20 &                             0.42 &              16 &              14 &                                  9 &                              22 &       1.20 \\
30 &       59 &                             0.18 &                             0.14 &                             0.40 &              42 &              78 &                                 17 &                              63 &       1.35 \\
31 &       57 &                             0.29 &                             0.21 &                             0.40 &              14 &              15 &                                 10 &                              24 &       1.06 \\
32 &       56 &                             0.18 &                             0.13 &                             0.38 &              35 &              23 &                                 43 &                              83 &       1.54 \\
33 &       69 &                             0.19 &                             0.10 &                             0.38 &              49 &              54 &                                 60 &                             108 &       2.11 \\
34 &       63 &                             0.21 &                             0.14 &                             0.39 &              23 &              20 &                                 22 &                              33 &       1.07 \\
35 &       69 &                             0.29 &                             0.22 &                             0.41 &              12 &               3 &                                 11 &                              28 &       1.01 \\
36 &       72 &                             0.24 &                             0.16 &                             0.36 &              21 &              22 &                                 16 &                              39 &       1.18 \\
37 &       63 &                             0.21 &                             0.16 &                             0.36 &              25 &              23 &                                 51 &                             142 &       1.11 \\
39 &       55 &                             0.27 &                             0.18 &                             0.36 &              11 &               3 &                                 12 &                              29 &       1.00 \\
40 &       54 &                             0.18 &                             0.10 &                             0.33 &              44 &              69 &                                 81 &                             151 &       1.39 \\
46 &       52 &                             0.08 &                             0.06 &                             0.30 &              57 &              73 &                                 59 &                             157 &       1.65 \\
48 &       53 &                             0.12 &                             0.08 &                             0.30 &              40 &             104 &                                 13 &                              44 &       1.82 \\
51 &       56 &                             0.19 &                             0.13 &                             0.29 &              12 &               5 &                                 17 &                              41 &       1.00 \\
58 &       58 &                             0.14 &                             0.09 &                             0.23 &              29 &              10 &                                 21 &                              66 &       1.00 \\
\bottomrule
\end{tabular}
\end{center}
\end{table*}

\section{Visualization of cluster data} 
\label{app:visualization}
The visualization of a cluster may give the impression of its dynamical state and set an expectation for the offset for both the readers and the person creating the figure. 
Being able to compute a scientifically accurate visualization, with well-justified parameter choices, is therefore very important in scientific literature.
For example, we inspected both the luminosity maps and the number density maps of the member galaxy populations and find that, with the same selection of bright galaxies of apparent $i$-band$ < 24.4$ at
$z=0.3$, the luminosity maps in general resemble the DM maps more closely than the number density maps. We argue that the KDE is more than a method for identifying the luminosity peaks.

In Figures \ref{fig:stamps1} and \ref{fig:stamps2}, we provide the visualization of a random selection of the Illustris clusters adopted in this work (in randomly chosen projections). Notations are as in Figure \ref{fig:select_peak_visualization}: for each cluster, we show the DM density distribution with very fine smoothing kernel (left) and the contours of the kernel density estimates (KDE) of the projected distribution of cluster galaxies obtained by weighting each galaxy by its i-band luminosity. For the results of Section \ref{sec:results}, we have actually used the DM density maps (and identified DM peaks) smoothed with 50-kpc kernels, however here we prefer to showcase the more highly resolved DM maps. At the following links, more Illustris cluster data visualization can be found, particularly:

\begin{itemize}
\item 2-kpc smooth DM and luminosity-weighted galaxy KDE maps: \href{http://goo.gl/WiDijQ}{http://goo.gl/WiDijQ} 
\item 50-kpc smooth DM and luminosity-weighted galaxy KDE maps: \href{http://goo.gl/89edcM}{http://goo.gl/89edcM}
\item DM maps of the most massive ~130 haloes in the Illustris volume: \href{https://goo.gl/kZUWrg}{https://goo.gl/kZUWrg}
\item KDE luminosity weighted galaxy maps of the most massive ~130 haloes in the Illustris volume: \href{https://goo.gl/R7VNi9}{https://goo.gl/R7VNi9}
\item KDE number density galaxy maps of the most massive ~130 haloes in the Illustris volume: \href{https://goo.gl/lmQUPd}{https://goo.gl/lmQUPd}.

\end{itemize}

\begin{figure*}
\begin{center}
	\includegraphics[width=0.98\linewidth]{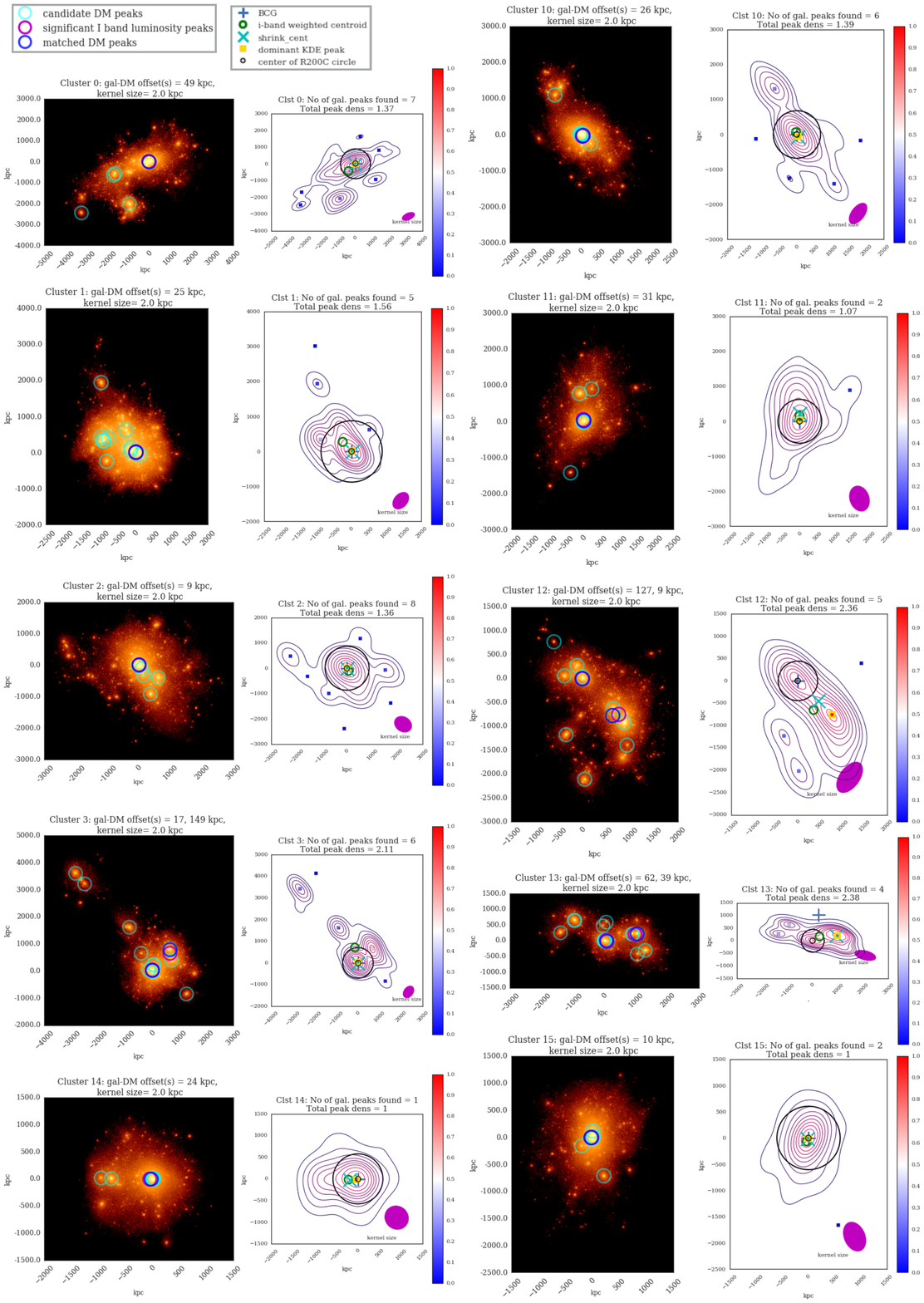}
	\caption{ Visualization of a random selection of clusters adopted in this work. Notations are as in Figure \ref{fig:select_peak_visualization}: for each cluster, we show the DM density distribution with very fine smoothing kernel (left panels) and the contours of the kernel density estimates (KDE) of the projected distribution of cluster galaxies obtained by weighting each galaxy by its i-band luminosity (right panels). The colorbars denote the relative density to the densest peak of the KDE luminosity peaks (square markers). For the results of Section \ref{sec:results}, we have actually used the DM density maps smoothed with 50-kpc kernels.
	   	\label{fig:stamps1}
	}
\end{center}	
\end{figure*}
\begin{figure*}
\begin{center}
	\includegraphics[width=0.98\linewidth]{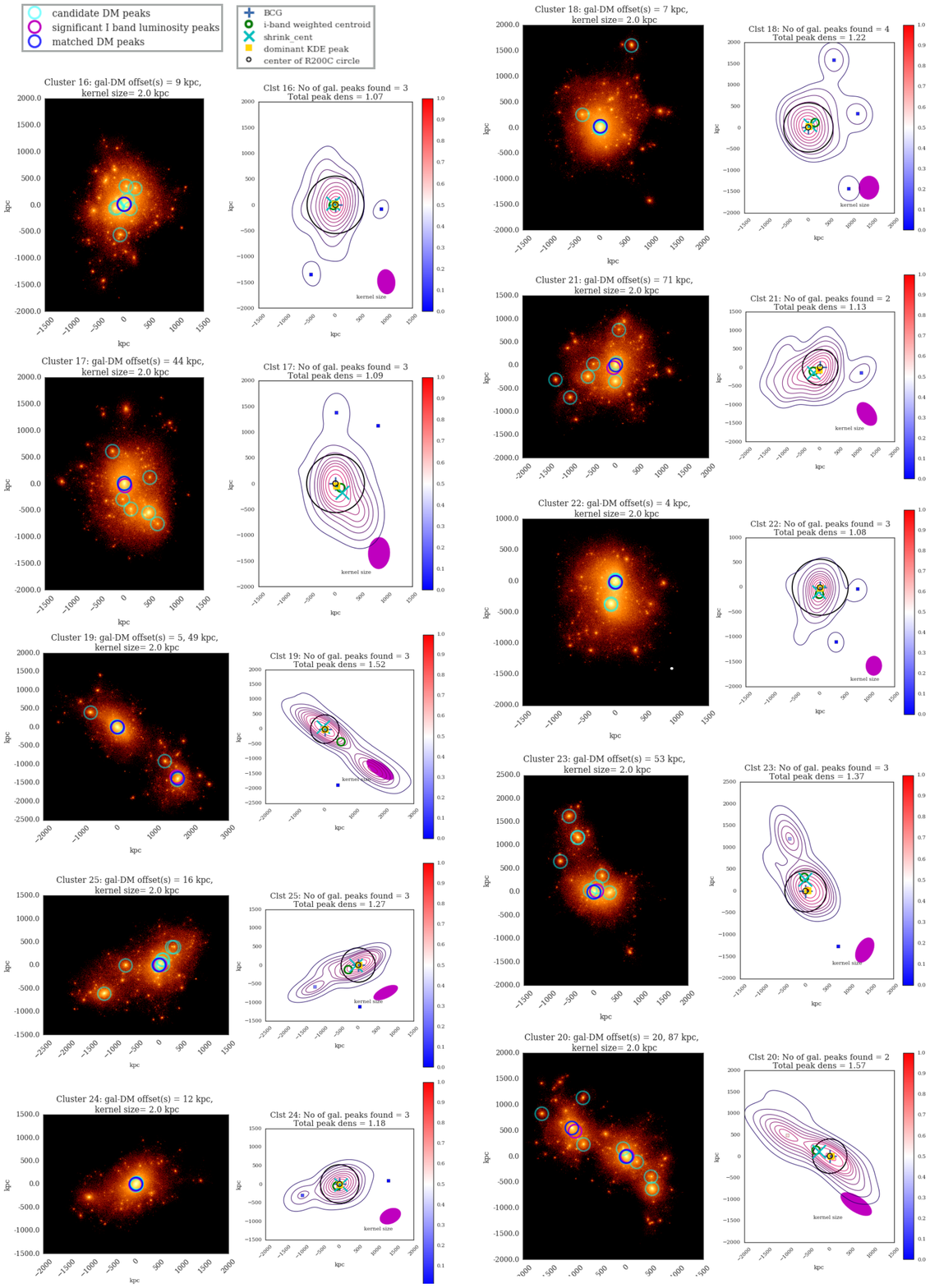}
	\caption{ As in Figure \ref{fig:stamps1}
	   	\label{fig:stamps2}.
	}
\end{center}	
\end{figure*}

\end{document}